\documentclass[english,a4paper]{article}
\usepackage[T1]{fontenc}
\usepackage[cp1250]{inputenc}
\usepackage{amssymb}
\usepackage{hyperref}
\usepackage{amsfonts}
\usepackage{graphicx}

\newtheorem{proposition}{Proposition}
\newtheorem{definition}{Definition}

\def\sizefig{0.5}

\begin{document}

\title{Geometric optimal control of the contrast imaging problem in Nuclear Magnetic Resonance}
\author{B. Bonnard\footnote{Institut de Math\'ematiques de Bourgogne, UMR CNRS 5584, 9 Avenue
        Alain Savary, BP 47 870 F-21078 DIJON Cedex FRANCE}, O.
        Cots, S. J. Glaser\footnote{Department of Chemistry, Technische Universit\"at M\"unchen, Lichtenbergstrasse 4, D-85747 Garching, Germany}, M. Lapert, D. Sugny\footnote{Laboratoire Interdisciplinaire Carnot de
Bourgogne (ICB), UMR 5209 CNRS-Universit\'e de Bourgogne, 9 Av. A.
Savary, BP 47 870, F-21078 DIJON Cedex, FRANCE,
dominique.sugny@u-bourgogne.fr} and Y. Zhang}

\maketitle

\begin{abstract}
The objective of this article is to introduce the tools to analyze
the contrast imaging problem in Nuclear Magnetic Resonance.
Optimal trajectories can be selected among extremal solutions of
the Pontryagin Maximum Principle applied to this Mayer type
optimal problem. Such trajectories are associated to the question
of extremizing the transfer time. Hence the optimal problem is
reduced to the analysis of the Hamiltonian dynamics related to
singular extremals and their optimality status. This is
illustrated by using the examples of cerebrospinal fluid / water
and grey / white matter of cerebrum.
\end{abstract}
\section{Introduction}
In a series of recent articles
\cite{assemat,BCS,energy,bonnardsugny,sugnykontz,zhang}, geometric
optimal control combined with adapted numerical schemes such as
the Hampath code \cite{hampath} is used to analyze the optimal
control of Kossakowsky-Lindblad equations
\cite{altafini,gorini,lindblad}. These equations describes the
evolution of a two-level dissipative quantum system whose dynamics
is governed by a three-dimensional system
\begin{eqnarray}\label{KL}
& & \frac{dx}{dt}=-\Gamma x+u_2z \nonumber\\
& & \frac{dy}{dt}=-\Gamma y-u_1z\\
& & \frac{dz}{dt}=\gamma_--\gamma_+z+u_1y-u_2x \nonumber,
\end{eqnarray}
the state variable $q=(x,y,z)$ belonging to the Bloch ball
$|q|\leq 1$ which is invariant for the dynamics since the
dissipative parameters $\Lambda=(\Gamma,\gamma_+,\gamma_-)$
satisfy $2\Gamma\geq \gamma_+\geq |\gamma_-|$. The control field
is $u=(u_1,u_2)$. The underlying optimal control problem consists
of minimizing the transfer time with a bound on the modulus of the
control or of minimizing the energy transfer $\int_0^T|u|^2dt$
with a fixed control duration.

Such a system is a model for the control of a molecule in a
dissipative environment using a laser field
\cite{mirrahimi,viellard} but also in Nuclear Magnetic Resonance
(NMR) spectroscopy where the dynamics of a spin 1/2 particle can
be described, up to a renormalization, by the Bloch equation which
is of the form (\ref{KL}) with the restriction $\gamma_-=\gamma_+$
\cite{ernst,khaneja,spin}. This implies that in this model, the
equilibrium point of the free motion is the north pole $(0,0,1)$
of the Bloch ball.

In NMR, we also recall that the control is a
transverse radio-frequency magnetic field in the $(x,y)$- plane, a
constant magnetic field being applied in the $z$- direction. In
this domain, a striking application of geometric optimal control
was a gain of 60 \% in the control duration of the saturation of a
spin 1/2 particle \cite{lapert}. The saturation problem consists
in bringing the magnetization vector of the sample from the
equilibrium point to the center of the Bloch ball
\cite{contrast1}. Such a control can be achieved by a standard NMR
technique, the inversion recovery sequence, composed of a bang arc
to invert the magnetization vector and a singular one along the
vertical $z$- axis to reach the target state. It can be shown that
the geometric time-optimal solution is the concatenation of a
bang, a horizontal singular arc, a bang and a final vertical
singular arc. The gain in the control duration has been shown
experimentally in \cite{lapert}. The experiments were performed
using the proton spins of H$_2$O in an organic solvent at room
temperature. This result shows that the optimized pulse sequence
can really be implemented with modern NMR spectrometers and a
reasonable match between theory and experiments.

Also this result is crucial because it confirms the ubiquity of
singular trajectories in the optimal control of nonlinear systems
\cite{bonnardchyba}. In the preceding example, contrary to the
apparent simplicity of the equations, the physical situation is
non trivial due to the two singular directions which are necessary
to compute the optimal solution. A direct generalization of this
problem is the one of the contrast in NMR imaging. The model is
obtained by considering two uncoupled spins, each of them being
solution of the Bloch equations (\ref{KL}) with different damping
coefficients $\Lambda_1=(\Gamma_1,\gamma_1)$,
$\Lambda_2=(\Gamma_2,\gamma_2)$, but controlled by the same
magnetic field. Denoting each system by
$$
\frac{dq_i}{dt}=F_i(q_i,\Lambda_i,u)
$$
where $q_i=(x_i,y_i,z_i)$ is the magnetization vector of each spin
particle, this leads to a system written shortly as
$$
\frac{dx}{dt}=F(x,u)
$$
where $x=(q_1,q_2)$. The associated optimal control problem is the
following: Starting from the equilibrium point of the dynamics
$x_0=((0,0,1),(0,0,1))$, the goal is to reach in a given transfer
time $T$ (which can be fixed or not), the final state $q_1(T)=0$
for the first spin while maximizing a cost $C(q_2(T))$ (e.g.
$|q_2(T)|^2$ or the projection of $q_2(T)$ on one axis). A subcase
of this problem is to restrict the system to $x_1=x_2=0$ by
considering only the component $u_1$ of the control field. Our aim
in this paper is to present a geometric study of this control
problem based on the analysis of the Hamiltonian dynamics given by
the Pontryagin Maximum Principle (PMP) \cite{pont}, the optimal control problem being a standard Mayer problem.

An important point in our analysis will be the introduction of
singular trajectories of the system $\frac{dx}{dt}=F(x,u)$ whose
control domain $N$ is a smooth submanifold of $\mathbb{R}^p$
defined as follows (one can assume that $N=\mathbb{R}^p$):
\begin{definition}
A control $u\in L^\infty ([0,T])$ is called singular on $[0,T]$ if
the derivative of the extremity mapping $E^{x_0,T}:~u\in
L^\infty\mapsto x(T,x_0,u)$, where $x(\cdot)$ denotes the response
to $u(\cdot)$ initiating from $x_0$ at $t=0$, is not of full rank.
\end{definition}
This definition is not the standard definition in the engineering litterature,
in particular it depends upon the control domain. But it is the correct
mathematical definition in optimal control since optimality is related to
openess properties of the extremity mapping.

A large amount of work has been done recently in control theory to
analyze the role of singular extremals. This can be summarized as
follows:
\begin{enumerate}
\item They are feedback invariant.
\item They can be computed using the PMP as solutions of
$$
\dot{x}=\frac{\partial H}{\partial p},~\dot{p}=-\frac{\partial
H}{\partial x},~\frac{\partial H}{\partial u}=0
$$
where $H(x,p,u))=\langle p,F(x,u)\rangle$ is the Hamiltonian lift
of the system.
\end{enumerate}
As such they are extremal solutions of any Mayer type problem
associated to a system where the cost and the boundary conditions
only give boundary conditions. Also recent works have shown how to
compute their first conjugate time, that is the first time such
that the extremity mapping becomes open. This time corresponds
also to the time where the trajectories lose their local
optimality. Theoretically, it is related to the concept of
singularity of Lagrangian manifolds \cite{bonnardchyba} and is
numerically implemented in the Hampath code \cite{hampath}.

Hence going back to the contrast imaging problem, a research
program is to analyze the Hamiltonian dynamics of the singular
extremals completed by numerical simulations to compute the
optimal solutions. This is a difficult task since the problem is
depending upon different relaxation parameters in the Bloch
equation. In this paper, we will present the geometric tools and
some preliminary numerical results in two particular cases by
considering only one component of the control field.

The organization of this article is the following. In the first
section, the Maximum Principle is introduced to select minimizers
among extremal solutions in a Mayer problem. The role of singular
extremals is presented and their optimality status is determined
using the concept of conjugate points. In a second section, a
thorough analysis of the geometric control of a single spin 1/2
particle is presented and it plays for specific values of the
parameters, an important role in the problem. In the final
section, we numerically analyze the geometry of singular extremals
in view of studying some specific cases in NMR. Numerical
computations of the optimal solution are also presented for two
regularized cost functionals.
\section{Geometric optimal control}\label{sec2}
\subsection{Preliminaries}
One considers a Mayer problem given by the following data~:
\begin{enumerate}
\item A smooth system $\frac{dx}{dt}=F(x,u)$, $x\in \mathbb{R}^n$
with fixed initial state $x_0$ and a transfer time $T$, the
controls being the set $\mathcal{U}=L^\infty ([0,T],U)$ of bounded measurable
mappings valued in a control domain $U \subset \mathbb{R}^p$.
\item A terminal manifold $M$ defined by $f(x)=0$ where
$f:~\mathbb{R}^n\to \mathbb{R}^k$ is a smooth mapping.
\item A cost to minimize : $\min_{u(\cdot) \in \mathcal{U}}C(q(T))$
where $C:~\mathbb{R}^n\to \mathbb{R}$ is a smooth regular mapping.
\end{enumerate}
The geometric setting is the following. Denote $x(t,u)$ the trajectory
initiating from $x_0$ and associated to $u$, $A(x_0,T)=\cup _{u\in\mathcal{U}}
x(T,u)$ the accessibility set at time $T$ and
introducing the manifold $C_m=\{f=0,C(x)=m\}$ where $m$ is a
parameter, an optimal control $u^*$ is such that
$x^*(T)=x(T,u^*)$ belongs to the boundary of $A(x_0,T)$,
$f(x^*(T,u^*))=0$ and $m$ is minimum.

\subsection{Pontryagin Maximum Principle}
The application of the maximum principle leads to the following
necessary conditions \cite{pont}.
\begin{proposition}
Let $u^*(\cdot)$ be an admissible control whose corresponding
trajectory $x^*(t)=x(t,u^*)$ is optimal. Then there exists an
absolutely continuous vector function $p^*(\cdot)$ and a scalar
$p_0\leq 0$ such that if we denote by $H$ the pseudo-Hamiltonian
$H(x,p,u)=\langle p,F(x,u)\rangle$, the following necessary
conditions are satisfied a.e. on $[0,T]$:
\begin{eqnarray}
& &\frac{dx^*}{dt}=\frac{\partial H}{\partial
p}(x^*,p^*,u^*),~\frac{dp^*}{dt}=-\frac{\partial H}{\partial
x}(x^*,p^*,u^*) \label{eq21}\\
& & H(x^*,p^*,u^*)=\max_{u\in U}H(x^*,p^*,u) \label{eq22}
\end{eqnarray}
together with the boundary conditions:
\begin{eqnarray}
& & f(x^*(T))=0 \label{eq23}\\
& & p^*(T)=p_0\frac{\partial C}{\partial x}(x^*(T))+\langle
\xi,\frac{\partial f}{\partial x}(x^*(T))\rangle ,\label{eq24}
\end{eqnarray}
$\xi\in \mathbb{R}^k$, $p_0\leq 0$ (transversality conditions).
\end{proposition}
\begin{definition}
We call extremals a triplet $(x,p,u)$ solution of (\ref{eq21}) and
of the maximization condition (\ref{eq22}). It is called a BC-
extremal if it satisfies the boundary conditions (\ref{eq23}) and
(\ref{eq24}).
\end{definition}
\subsection{A review of the properties of singular trajectories}
Next we present some concepts and properties about singular
trajectories which are important in our analysis, see
\cite{bonnardchyba} for a complete presentation.

We have the following characterization of singular control which
allows a practical computation.
\begin{proposition}
The control $u(\cdot)$ and the corresponding trajectory $x(\cdot)$
are singular on $[0,T]$ if and only if there exists a non zero
adjoint vector $p(\cdot)$ such that $(x,p,u)$ is solution a.e. on
$[0,T]$ of
\begin{equation}
\dot{x}=\frac{\partial H}{\partial p},~\dot{p}=-\frac{\partial
H}{\partial x},~\frac{\partial H}{\partial u}=0
\end{equation}
where $H(x,p,u)=\langle p,F(x,u)\rangle$ is the Hamiltonian lift.
Moreover for each $0<t\leq T$, $p(t)$ is orthogonal to
$\textrm{Im}E'^{x_0,t}(u|_{[0,t]})$.
\end{proposition}
\begin{definition}
A singular extremal is a triple $(x,p,u)$ solution of the above
equations. It is called:
\begin{enumerate}
\item Regular if $\frac{\partial^2 H}{\partial u^2}$ is of maximal rank.
\item Strongly normal if for each $0<t_1<t_2\leq T$,
$\textrm{Im}E'^{x(t_1),t_2-t_1}(u|_{[t_1,t_2]})$ is of corank one.
\item Exceptional if $H=0$.
\end{enumerate}
\end{definition}
\noindent\textbf{Computation in the regular case:} Using the
condition $\frac{\partial^2H}{\partial u^2}\neq 0$, one can solve
locally the equation $\frac{\partial H}{\partial u}=0$ and compute
the singular control as a function $\hat{u}(z)$, $z=(x,p)$ and
plugging such $\hat{u}$ in $H$ defines a true Hamiltonian denoted again
$H(z)$. If $\Pi$ is the standard projection $(x,p)\mapsto
x$, one can define the exponential mapping
$\exp_{x_0}:~(t,p)\mapsto \Pi(\exp[t\vec{H}(x_0,p)])$ where $x_0$
is fixed. This leads to the following definition.
\begin{definition}
Let $z(t)=(x(t),p(t))$ be the reference extremal solution of
$\vec{H}$. The time $t_c$ is said to be \emph{geometrically
conjugate} if $\exp_{x_0}$ is not of maximal rank at $(t_c,p(0))$.
\end{definition}
We have the following standard test:
\begin{proposition}
The time $t_c$ is geometrically conjugate if and only if there
exists a non trivial Jacobi field $J(t)$ solution of the
variational equation $\delta \dot{z}=d\vec{H}(z(t))\delta z$ and
vertical at time 0 and $t_c$: $d\Pi(J(0))=d\Pi(J(t_c))=0$.
\end{proposition}
The following result is crucial in our optimality analysis:
\begin{proposition}
In the strongly normal case and in the non exceptional situation,
the extremity mapping $E^{x_0,T}$ is open for the $L^\infty$-
topology at $u_{|[0,t]}$ where $t>t_{1c}$.
\end{proposition}
\textbf{Application:} One consider a control system of the form
$F(x,u)=F_0(x)+u_1F_1(x)+u_2F_2(x)$ where the control domain $U$
is the disk $u_1^2+u_2^2\leq 1$. The Hamiltonian is
$H=H_0+u_1H_1+u_2H_2$ where $H_i=\langle p,F_i(x)\rangle$. The
maximization condition (\ref{eq22}) leads to
\begin{equation}\label{eq27}
u_i=\frac{H_i}{\sqrt{H_1^2+H_2^2}},~i=1,2
\end{equation}
outside the switching surface $\Sigma$: $H_1=H_2=0$. The
corresponding extremals are called of order zero and there are
solutions of the smooth vector field defined by
$H(z)=H_0+\sqrt{H_1^2+H_2^2}$. The corresponding solutions are
regular singular extremals if one restricts the control domain to
the unit sphere $S^1$. Introducing $u_1=\cos\alpha$,
$u_2=\sin\alpha$ and extending the system using $\dot{\alpha}=v$,
they correspond to singular trajectories of the extended system:
$$
\dot{x}=F_0+\cos\alpha F_1+\sin \alpha F_2,~\dot{\alpha}=v.
$$
\textbf{The case of affine systems:} For optimality analysis, one
restricts our study to a single input affine system:
$\dot{x}=F_0+u_1F_1$, $|u_1|\leq 1$. Relaxing the control bound,
singular trajectories are parameterized by the constrained
Hamiltonian system:
$$
\dot{x}=\frac{\partial H}{\partial p},~\dot{p}=-\frac{\partial
H}{\partial x},~\frac{\partial H}{\partial u_1}=H_1=0.
$$
The singular extremals are not regular and the constraint $H_1=0$
has to be differentiated along an extremal to compute the controls. Introducing the
Lie brackets of two vector fields $X$, $Y$ computed with the
convention
$$
[X,Y](x)=\frac{\partial X}{\partial x}(x)Y(x)-\frac{\partial
Y}{\partial x}(x)X(x),
$$
and related to the Poisson bracket of the Hamiltonian lifts $H_X$,
$H_Y$ by the rule $\{H_X,H_Y\}=H_{[X,Y]}$, one gets:
$$
H_1=\{H_1,H_0\}=\{\{H_1,H_0\},H_0\}+u_1\{\{H_1,H_0\},H_1\}=0.
$$
A singular extremal such that $\{\{H_1,H_0\},H_1\}\neq 0$ is
called of minimal order and the corresponding control is given by
\begin{equation}\label{eq28}
u_{1s}=-\frac{\{\{H_1,H_0\},H_0\}}{\{\{H_1,H_0\},H_1\}}.
\end{equation}
Plugging such $u_{1s}$ into $H$ defined a true Hamiltonian, whose
solutions initiating from $H_1=\{H_1,H_0\}=0$ defined the singular
extremals of order zero. They are related to the regular case
using the following Goh transformation. Assuming $F_1$ non zero,
then there exists a coordinate system $(x_1,x_2,\cdots,x_n)$ on an
open set $V$ such that $F_1=\frac{\partial}{\partial x_n}$ and the
system splits into:
$$
\dot{x}'=F'(x',x_n),~\dot{x}_n=F_0'(x')+u_1
$$
where $x'=(x_1,\cdots,x_{n-1})$ and the system $F'$ defined on an
open subset $V'$ where $x_n$ is taken as the control variable is
called the reduced system. We introduce the reduced Hamiltonian
$H'(x',p',x_n)=\langle p',F'(x',x_n)\rangle$. One has:
\begin{eqnarray}
& &\frac{\partial}{\partial t}\frac{\partial H}{\partial
u}=\{H_1,H_0\}=-\frac{\partial H'}{\partial x_n} \label{eq29} \\
& & \frac{\partial}{\partial u}\frac{\partial^2}{\partial
t^2}\frac{\partial H}{\partial
u}=\{\{H_1,H_0\},H_1\}=-\frac{\partial ^2H'}{\partial x_n^2}
\label{eq210}.
\end{eqnarray}
This gives the relation between the affine singular case and the
regular one.
\subsection{High-order maximum principle in the affine case}
As a consequence and using the generalized Legendre Clebsch
condition deduced from the high-order maximum principle
\cite{krener}, one gets the following.

Consider the Mayer problem for an affine system of the form
$\dot{x}=F_0(x)+u_1F_1(x)$, $|u_1|\leq 1$. Then the following
conditions are necessary for optimality:
\begin{eqnarray*}
& & \dot{x}=\frac{\partial H}{\partial p},~\dot{p}=-\frac{\partial
H}{\partial x} \\
& & H(x,p,u)=\max_{|v|\leq 1}H(x,p,v)
\end{eqnarray*}
with the boundary conditions
$$
f(x(T))=0,~p(T)=p_0\frac{\partial C}{\partial x}+\langle
\xi,\frac{\partial f}{\partial x}\rangle,~p_0\leq 0.
$$
Moreover if the control is singular and non saturating, i.e.
$|u_{1s}|<1$, the generalized Legendre-Clebsch condition must
hold:
\begin{equation}
\{\{H_1,H_0\},H_1\}\geq 0 .\label{eq211}
\end{equation}
\subsection{Generic classification of the bang-bang extremals near
the switching surface} An important issue in the contrast problem
is to apply the results from \cite{kupka} to classify the extremal
curves near the switching surface. The switching surface is the
set $\Sigma:~H_1=0$, while the switching function is $t\mapsto
\Phi(z(t))=H_1(z(t))$, where $z(t)$ is an extremal curve. Let
$\Sigma_s:~H_1=0=\{H_1,H_0\}$. The singular extremals are entirely
contained in $\Sigma_s$. A bang-bang extremal $z(t)$ on $[0,T]$ is
an extremal curve with a finite number of switching times $0\leq
t_1<\cdots <t_n\leq T$. We denote by $\xi_+$, $\xi_-$ the regular
arcs for which $u=\pm 1$ and by $\xi_s$ a singular arc;
$\xi_1\xi_2$ denotes an arc $\xi_1$ followed by an arc
$\xi_2$.\\
\textbf{Ordinary switching time.} It is a time $t$ such that a
bang-bang arc switches with the condition $\Phi(t)=0$ and
$\dot{\Phi}(t)=\{H_1,H_0\}\neq 0$. According to the maximum
principle near $\Sigma$, the extremal is of the form $\xi_-\xi_+$
if $\dot{\Phi}(t)>0$ and $\xi_+\xi_-$ if
$\dot{\Phi}(t)<0$.\\
\textbf{Fold case.} It is the case where a bang arc has a contact
of order 2 with the switching surface. Denoting
$\ddot{\Phi}_\pm=\{\{H_1,H_0\},H_0\}\pm \{\{H_1,H_0\},H_1\}$ the
second derivative of the switching function, if non zero, we have
three cases:
\begin{enumerate}
\item \emph{Hyperbolic case:} At the switching point, one has
$\ddot{\Phi}_+>0$ and $\ddot{\Phi}_-<0$. At $\Sigma_s$, a
connection is possible with a singular extremal which is strictly
admissible and satisfies the strong Legendre-Clebsch condition.
The extremals are bang-singular-bang $\xi_\pm\xi_s\xi_\pm$.
\item \emph{Elliptic case:} At the switching point, one has
$\ddot{\Phi}_+<0$ and $\ddot{\Phi}_->0$. A connection with the
singular extremal is not possible and every extremal curve is
bang-bang but with no uniform bound on the number of switchings.
\item \emph{Parabolic case:} It is the situation where
$\ddot{\Phi}_+$ and $\ddot{\Phi}_-$ have the same sign at the
switching point. One can check that the singular extremal is not
admissible and every extremal curve near the switching point is
bang-bang with at most two switchings, i.e. $\xi_+\xi_-\xi_+$ or
$\xi_-\xi_+\xi_-$.
\end{enumerate}
\subsection{The concept of conjugate points in the affine
case}\label{secconj} According to \cite{bonnardchyba}, this
concept is related to the notion of conjugate points in the
regular case using the Goh reduction. The important property is
the following geometric characterization. Let
$z(\cdot)=(x(\cdot),p(\cdot))$ be a singular extremal associated
to the control defined by (\ref{eq28}). Assuming that it is
strictly admissible, one can embed the singular extremal into a
surface $S$ formed by all the singular extremals starting from
$x_0=\Pi(z(0))$ and with initial adjoint vector $p$ such that
$|p-p(0)|\leq \varepsilon$. Up to the first conjugate point, the
extremal synthesis is bang-singular-bang $\xi_\pm\xi_s\xi_\pm$,
where bang arcs will be in the neighborhood of the reference
singular extremal, which is related to the problem of extremizing
the transfer time and hence to the Mayer problem. This synthesis
is also valid in a $C^0$- neighborhood of the reference case in
the limit situation where $m\to +\infty $, $m$ being the control
bound.
\subsection{Application to the contrast problem}
A direct application is the contrast problem with the boundary
condition $q_1(T)=0$ and the cost $|q_2(T)|^2$. Splitting the
adjoint vector into $p=(p_1,p_2)$, we deduce the transversality
condition $p_2(T)=-2p_0q_2(T)$, $p_0\leq 0$. The case $p_0=0$
gives $p_2(T)=0$. Since the system splits into:
$$
\dot{q}_1=F_1'(q_1,u),~\dot{q}_2=F_2'(q_2,u)
$$
the adjoint system decomposes into:
$$
\dot{p}_1=-p_1\frac{\partial F_1'}{\partial
q_1},~\dot{p}_2=-p_2\frac{\partial F_2'}{\partial q_2}
$$
where $p=(p_1,p_2)$ is written as a row vector. The condition
$p_2(T)=0$ corresponds to a second spin which is not controlled.
In the non trivial case, $p_0$ is non zero and it can be
normalized to $p_0=-1/2$.
\subsection{The embedding results}
From the previous results, one deduces the following propositions.
\begin{proposition}
The time minimizing solutions of the first spin 1/2 particle can
be embedded as extremals of the contrast problem, with $p_0=0$.
\end{proposition}
\begin{proposition}
In the contrast problem, the extremals of the single-input case
are extremals of the bi-input case.
\end{proposition}
\section{The single spin 1/2 case}\label{sec3}
Since in the contrast problem, the magnetization vector of the
first particle has to be set to 0, an important issue is to
analyze this task and the underlying problem of reaching this
target in minimum time. Besides, the optimal solutions of such a
problem can be embedded into the extremal solutions of the
contrast problem. Indeed, if the transfer time in the contrast
problem is exactly this minimum time, they are the only solutions
satisfying the boundary conditions. Hence, in this section, based
on the preliminary work \cite{lapert}, we make a thorough analysis
of the single input case, with an emphasis put on the role of
singular trajectories.
\subsection{Preliminaries}
First of all, since the initial condition is on the $z$- axis of
revolution of the system, the control problem can be restricted to
the 2D- meridian of the Bloch ball and the control field reduced
to only one component \cite{BCS,bonnardsugny}. The system is
$\dot{q}=F_0(q)+u_1F_1(q)$, $|u_1|\leq m$, where
\begin{eqnarray*}
& & F_0=-\Gamma y\frac{\partial}{\partial y}+\gamma
(1-z)\frac{\partial}{\partial z} \\
& & F_1=-z\frac{\partial}{\partial y}+y\frac{\partial}{\partial z}
.
\end{eqnarray*}
Denoting $\delta=\gamma-\Gamma$, the following Lie brackets are
relevant in our analysis:
\begin{eqnarray*}
& &[F_1,F_0]=(-\gamma+\delta z)\frac{\partial}{\partial y}+\delta
y\frac{\partial}{\partial z}\\
& &
[[F_1,F_0],F_0]=(\gamma(\gamma-2\Gamma)-\delta^2z)\frac{\partial}{\partial
y}+\delta^2y\frac{\partial}{\partial z} \\
& & [[F_1,F_0],F_1]=2\delta y\frac{\partial}{\partial
y}+(\gamma-2\delta z)\frac{\partial}{\partial z}.
\end{eqnarray*}
\subsection{Singular trajectories and optimality}
The singular trajectories are located on the set
$S:~\textrm{det}(F_1,[F_1,F_0])=0$, which is given in our case by
$y(-2\delta z+\gamma)=0$. Hence it is formed by the $z$- axis of
revolution $y=0$ and the horizontal direction
$z=\gamma/(2\delta)$. The singular control is given by
$D'+u_{1s}D=0$ where $D=\textrm{det}(F_1,[[F_1,F_0],F_1])$ and
$D'=\textrm{det}(F_1,[[F_1,F_0],F_0])$.
\begin{itemize}
\item For $y=0$, one has $D=-z(\gamma-2\delta z)$ and $D'=0$. The
singular control is zero and the singular arc is solution of
$$
\dot{y}=-y,~\dot{z}=\gamma(1-z)
$$
where the equilibrium point $(0,1)$ is stable if $\gamma\neq 0$.
\item For $z=\gamma/(2\delta)$, $D=-2\delta y^2$, $D'=y\gamma
(2\Gamma-\gamma)$ and $u_{1s}=\gamma(2\Gamma-\gamma)/(2\delta y)$,
$2\Gamma-\gamma\geq 0$. Hence along the horizontal direction, the
flow is
$$
\dot{y}=-\Gamma y-\frac{\gamma^2(2\Gamma-\gamma)}{4\delta^2 y},
$$
and $|u_{1s}|\to +\infty$ when $y\to 0$.
\end{itemize}
More precisely, along the horizontal singular line, the following
proposition is crucial.
\begin{proposition}
If $\gamma\neq 0$, the singular control along the horizontal
singular line is in the $L^1$ but not in the $L^2$- category, near
$y=0$.
\end{proposition}
This can be straightforwardly shown by using the relations:
\begin{equation}\label{div}
\int_{t_0}^{t_1}u_{1s}(t)^2dt=\int_{y_0}^0u_{1s}(y)\frac{dy}{\dot{y}}
\end{equation}
where $t_0$ and $t_1$ are the initial and final times along the
singular arc and $y_0$ the initial $y$- coordinate of this arc.
One deduces that the integrand of Eq. (\ref{div}) scales as $1/y$
when $y\to 0$ and that the corresponding integral has a
logarithmic divergence.

In order to study the optimality of the singular directions, one
uses the generalized Legendre-Clebsch condition, which takes the
following form for a 2D-system. Let $D''=\det (F_1,F_0)=\gamma
z(z-1)+\Gamma y^2$. The set $C=\{D''=0\}$ is the collinear set. If
$\gamma\neq 0$, this set is not reduced to a point and the
intersection with the horizontal singular line is empty, except in
the case $\gamma=2\Gamma$. Singular lines are fast if $DD''>0$ and
slow if $DD''<0$.

To complete the optimality analysis, one introduces the clock form
$\omega=pdq$ which is defined outside the collinearity set $C$ by
the relations $\langle p,F_0\rangle=1$ and $\langle
p,F_1\rangle=0$, the sign of $d\omega$ being given by
$y(\gamma-2\delta z)$.This form allows to deduce the optimality of
singular extremals and to compare two different regular extremals
when they do not cross the singular and collinearity sets
\cite{bonnardchyba}.\\
\textbf{Parameters conditions}\\
The interesting case is when the horizontal singular line
$z_0=\gamma/(2\delta)$ cuts the Bloch ball $|q|\leq 1$, which
gives the condition $\Gamma>3\gamma/2$ and $-1<z_0<0$. Using the
generalized Legendre-Clebsch condition, one deduces that the
horizontal line is optimal and the $z$- axis of revolution is
optimal if $1>z>z_0$. In particular, this line is slow in the
domain $z_0>z>-1$. Using the clock form, one can deduce that near
the origin, the broken singular arc formed by a horizontal arc
followed by a vertical line is time-minimal for the unbounded
case, provided admissible controls are extended to $L^1$.
 Note also that such a broken singular trajectory is not
in $L^2$ and is not optimal for the energy
minimization problem \cite{energy}.

Having made this optimality analysis, one can deduce the time
minimal optimal synthesis near the origin which is introduced
next.
\subsection{The SiSi singularity (Interaction between two singular arcs)}
\begin{figure}
\centering
\includegraphics[width=\sizefig\textwidth]{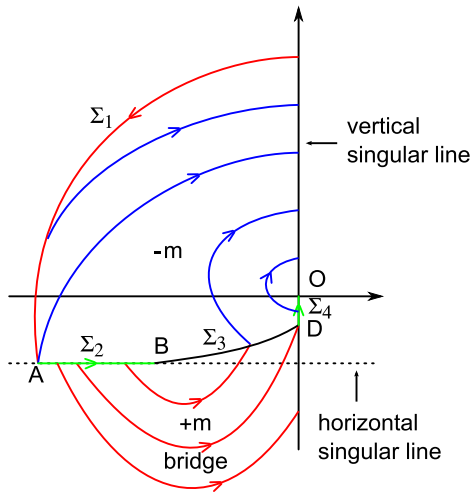}
\caption{\label{fig1} Schematic representation of the optimal
synthesis when the initial point of the dynamics is the north
pole. An arbitrary zoom has been used to construct the figure.
Regular curves are plotted in blue (dark) and red (dark grey) for
control fields equal to $-m$ and $+m$, respectively. The optimal
singular trajectories are displayed in green (light grey). The
black line is the switching curve, while the dashed one is the
non-admissible part of the horizontal singular line.}
\end{figure}

Assume $\gamma\neq 0$, $\Gamma>\frac{3}{2}\gamma$, $|u_1|\leq m$
and the control bound is large enough such that the bang arc
$u_1=m$ starting from the north pole $(0,1)$ intersects the
horizontal singular arc $z_0=\gamma/(2\delta)$ at a point A. The
horizontal singular line is admissible up to a saturating point B.
The time minimal synthesis, with initial point A, is represented
on Fig. \ref{fig1}.
Due to the saturation phenomenon at B, there is a birth of a
switching locus $\Sigma_3$, but the remarkable fact due to the
interaction between the horizontal and the vertical fast singular
directions is the following concept.
\begin{definition}
We call bridge between the horizontal singular arc and the
vertical singular one, the bang arc, such that the concatenation
singular-bang-singular is optimal.
\end{definition}
This concept is important and leads to a generalization in higher
dimension, which plays an important role in the contrast problem.

In order to compute the global optimal synthesis provided $m$ is
large enough, we must analyze the synthesis near the north pole,
which is presented next.

\subsection{The SiCo singularity (Interaction between the collinear
set and the singular set)}
Observe that the north pole is a stable fixed point for the free
motion and the vertical singular direction is a fast direction,
near the north pole, provided $z<1$, as a consequence of the
strong Legendre-Clebsch condition. The collinear set $C$
corresponds to an oval below the line $z=1$. Using polar
coordinates $y=r\sin\phi$, $z=r\cos\phi$, one gets
$$
r\frac{dr}{dt}=-\Gamma y^2-\gamma (z-1)=-D''.
$$
Hence, $r$ which represents the purity of the quantum system
decreases outside the oval and increases inside. The north pole is
a singularity of the optimal problem which combines a collinear
situation with a singular one, but the analysis of the
time-minimal synthesis near this point is simple because of the
symmetry of revolution. Indeed, the only way to leave this
singularity is to use a bang arc $u=\pm m$, which gives a boundary
arc of the accessibility set. This first bang arc is followed by
another bang arc $u=\mp m$ to fill the interior of the domain and
to reach the vertical singular axis. This gives an optimal
bang-bang policy (see the top part of Fig. \ref{fig1}).
\subsection{The global synthesis}
Under our assumptions ($\Gamma>3\gamma/2>0$, $m$ large enough),
the global time minimal synthesis starting from the north pole is
easily obtained gluing the two previous syntheses (the one
associated to the SiSi case with the one of the SiCo case). It is
represented on Fig. \ref{fig1}. The switching locus is formed by
the arc starting from the north pole and reaching the horizontal
singular arc at A (it is denoted $\Sigma_1$ in the figure), the
horizontal singular segment $\Sigma_2$ between the points A and B,
the switching locus $\Sigma_3$ due to the saturation phenomenon
and the part of the vertical singular direction between D and 0
(the $\Sigma_4$ segment), D being the extremity of the bridge. The
bang arc with $u=-m$ starting from A is separating the two
domains, one with a bang-bang policy and the other containing a
non trivial singular arc.

At the limit, when $m\to +\infty$, it gives the synthesis
constructed in Ref. \cite{lapert} where the total time to reach
the origin is formed by the time to follow the broken
singular-singular arc between A and 0. Observe also that according
to our analysis, the usual policy in NMR, the inversion recovery
sequence, where only the vertical singular arc is used, is slow if $z<z_0$.

Also, note that the switching locus has a complicated structure,
but due to the symmetry of revolution, all the cut points, i.e.
the first points where the extremal trajectories cease to be
optimal, are on the vertical $z$- axis where two symmetric
solutions starting respectively on the left and right part of the
Bloch disk intersect at the same time.
\section{Preliminary results in the contrast problem}
As mentioned in the introduction, the goal of the contrast problem
is to bring the magnetization vector of spin 1 towards the center
of the Bloch ball together maximizing the modulus of the
magnetization vector of the other specie. Note that such a
computation could have potential applications in magnetic
resonance imaging in order to optimize the contrast of a given
imaging \cite{contrast1,contrast2}. Roughly speaking, the species
with a zero magnetization will appear dark, while the other
species with a maximum modulus of the magnetization vector will be
white. We introduce in the following a simple model reproducing
the main features of this control problem. We describe the general
structure of the optimal solution and we compute them for two
particular examples.
\subsection{The model system}
Each spin 1/2 particle is governed by the Bloch equation:
\begin{eqnarray*}
& &\frac{dM_x}{dt}=-\frac{M_x}{T_2}+\omega_y M_z \\
& &\frac{dM_y}{dt}=-\frac{M_y}{T_2}-\omega_x M_z \\
& &\frac{dM_z}{dt}=\frac{(M_0-M_z)}{T_1}+\omega_xM_y-\omega_yM_x
\end{eqnarray*}
where the state variable is the magnetization vector and $T_1$,
$T_2$ are the relaxation times. The control is the magnetic field
$\omega=(\omega_x,\omega_y)$ which is bounded by $|\omega|\leq
\omega_{max}$. We use the normalization introduced in
\cite{lapert}. The normalized coordinates are
$q=(x,y,z)=(M_x,M_y,M_z)/M_0$. In these coordinates, the
equilibrium point is the north pole $(0,0,1)$ and the normalized
control is
$u=(u_x,u_y)=\frac{2\pi}{\omega_{max}}(\omega_x,\omega_y),~|u|\leq
2\pi$, while the normalized time is given by
$\tau=\omega_{max}t/(2\pi)$. Hence the system takes the form:
\begin{eqnarray*}
& & \dot{x}=-\Gamma x+u_y z\\
& & \dot{y}=-\Gamma y-u_x z\\
& & \dot{z}=\gamma (1-z)+(u_xy-u_yx)
\end{eqnarray*}
where $\Gamma=2\pi/(\omega_{max}T_2)$ and
$\gamma=2\pi/(\omega_{max}T_1)$. In the experiments,
$\omega_{max}$ can be chosen up to 15 000 Hz but the value
$2\pi\times 32.3$ Hz will be considered in this paper. The
experiments are done for the
 contrast problems of the cerebrospinal fluid/water \cite{water} and the grey/white matter of cerebrum cases \cite{matter}. In
 the cerebrospinal fluid/water situation, the relaxation parameters for the
 first spin describing the fluid are $T_1=2000$ ms and $T_2=200$ ms, while for the
 second spin $T_1=T_2=2500$ ms. In the second example, the rates of the grey
 matter are taken to be $T_1=920$ ms and $T_2=100$ ms, the rates for the white
 matter being $T_1=780$ ms and $T_2=90$ ms.
\subsection{Computation of the singular flow}
One restricts to the situation where the control field has only
one component and the contrast problem is governed by the
differential system $\dot{q}=F_0(q)+u_1F_1(q)$,
$q=(y_1,z_1,y_2,z_2)$:
\begin{eqnarray*}
& & F_0=\sum _{i=1}^2[-\Gamma_i y_i\frac{\partial}{\partial
y_i}+\gamma_i(1-z_i)\frac{\partial}{\partial z_i}] \\
& & F_1=\sum_{i=1}^2(-z_i\frac{\partial}{\partial
y_i}+y_i\frac{\partial}{\partial z_i}) .
\end{eqnarray*}
Denoting $\delta_i=\gamma_i-\Gamma_i$, $i=1,2$ one has:
\begin{eqnarray*}
& &
[F_1,F_0]=\sum_{i=1}^2(-\gamma_i+\delta_iz_i)\frac{\partial}{\partial
y_i}+\delta_iy_i\frac{\partial}{\partial z_i} \\
& &
[[F_1,F_0,F_0]]=\sum_{i=1}^2[\gamma_i(\gamma_i-2\Gamma_i)-\delta_i^2z_i]\frac{\partial}{\partial
y_i}+\delta_i^2y_i\frac{\partial}{\partial z_i} \\
& &
[[F_1,F_0],F_0]=\sum_{i=1}^22\delta_iy_i\frac{\partial}{\partial
y_i}+(\gamma_i-2\delta_iz_i)\frac{\partial}{\partial z_i}
\end{eqnarray*}
and the corresponding singular flow is defined by:
$$
H_{1}=\{H_{1},H_{0}\}=\{\{H_{1},H_{0}\},H_{0}\}+u_{1s}\{\{H_{1},H_{0}\},H_{1}\}=0.
$$
Since the equations are linear with respect to $p$, for each
initial condition $q_0$, this defines a two-dimensional surface
$S(q_0)$ in the state space. An additional condition is provided
by the generalized Legendre-Clebsch condition:
$\{\{H_1,H_0\},H_1\}\geq 0$. The structure of this surface is
related to the relaxation parameters $(\Gamma_i,\gamma_i)$.

If the transfer time is not fixed, this leads to the additional
constraints $H_0=0$. In this case, the singular flow defines a
single vector field in the state space, since the adjoint vector
can be eliminated and the restricted singular control is given by:
$$
u_{1s}=-\frac{D'(q)}{D(q)}
$$
where
\begin{eqnarray*}
& & D'(q)=\textrm{det}(F_0,F_1,[F_1,F_0],[[F_1,F_0],F_0]) \\
& & D(q)=\textrm{det}(F_0,F_1,[F_1,F_0],[[F_1,F_0],F_1])
\end{eqnarray*}
with the corresponding vector field
$$
\frac{dq}{dt}=F_0(q)-\frac{D'(q)}{D(q)}F_1(q)
$$
which can be analyzed using the time reparameterization
$d\tau=dt/D(q(\tau))$. In this framework, singular trajectories
are used to classify the systems.

In the general case, a similar computation shows that the singular
trajectories are solutions of an equation of the form:
\begin{equation}\label{eqsing}
\frac{dq}{dt}=F_0(q)-\frac{D'(q,\lambda)}{D(q,\lambda)}F_1(q)
\end{equation}
where $\lambda$ is a one dimensional time dependent parameter
whose dynamics is deduced from the adjoint equation. The solutions
of Eq. (\ref{eqsing}) emanating from $q_0$ will form $S(q_0)$.
\subsection{Numerical simulations on singular trajectories} 
\begin{figure}
\centering
\includegraphics[width=\sizefig\textwidth]{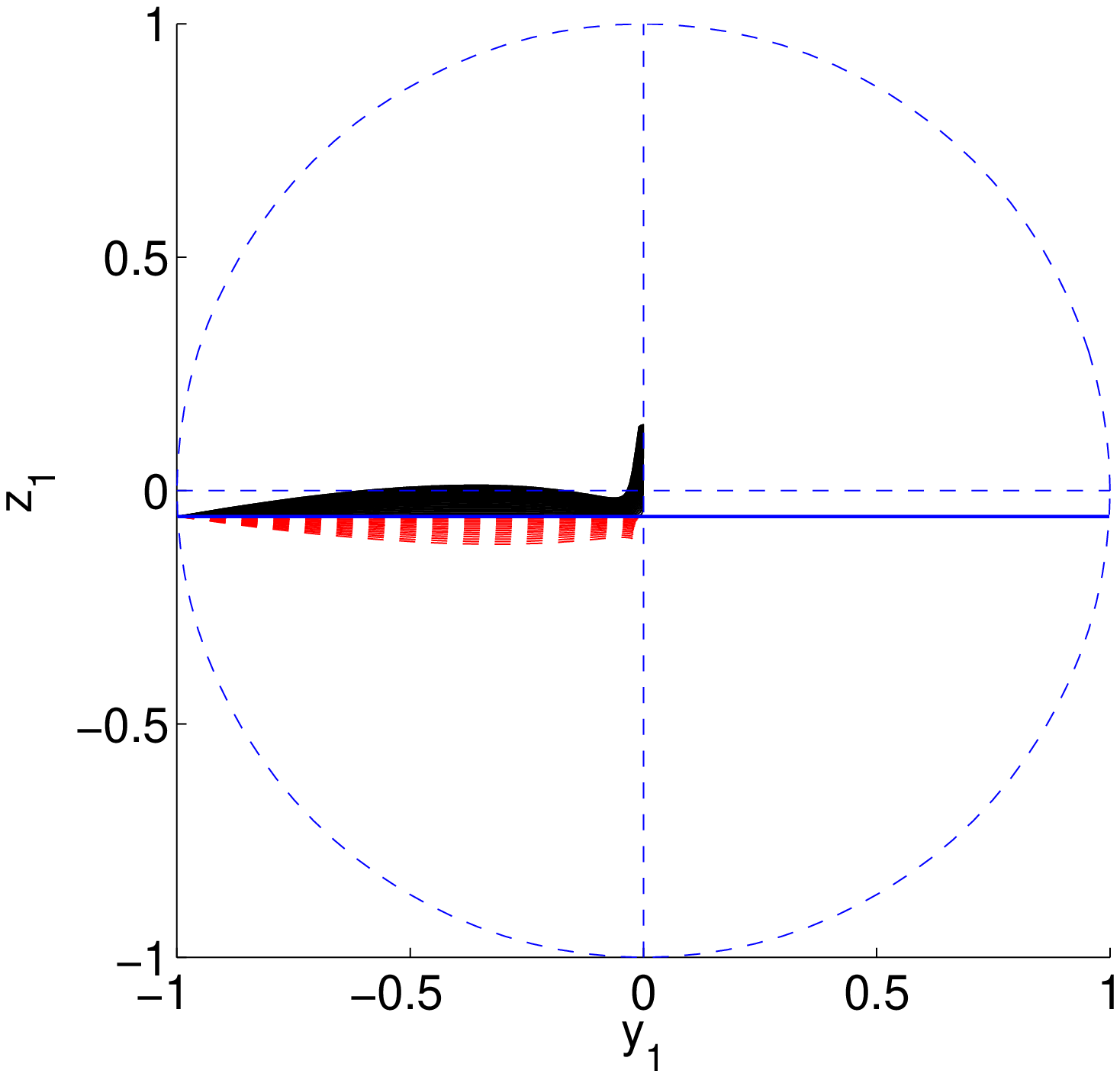}
\includegraphics[width=\sizefig\textwidth]{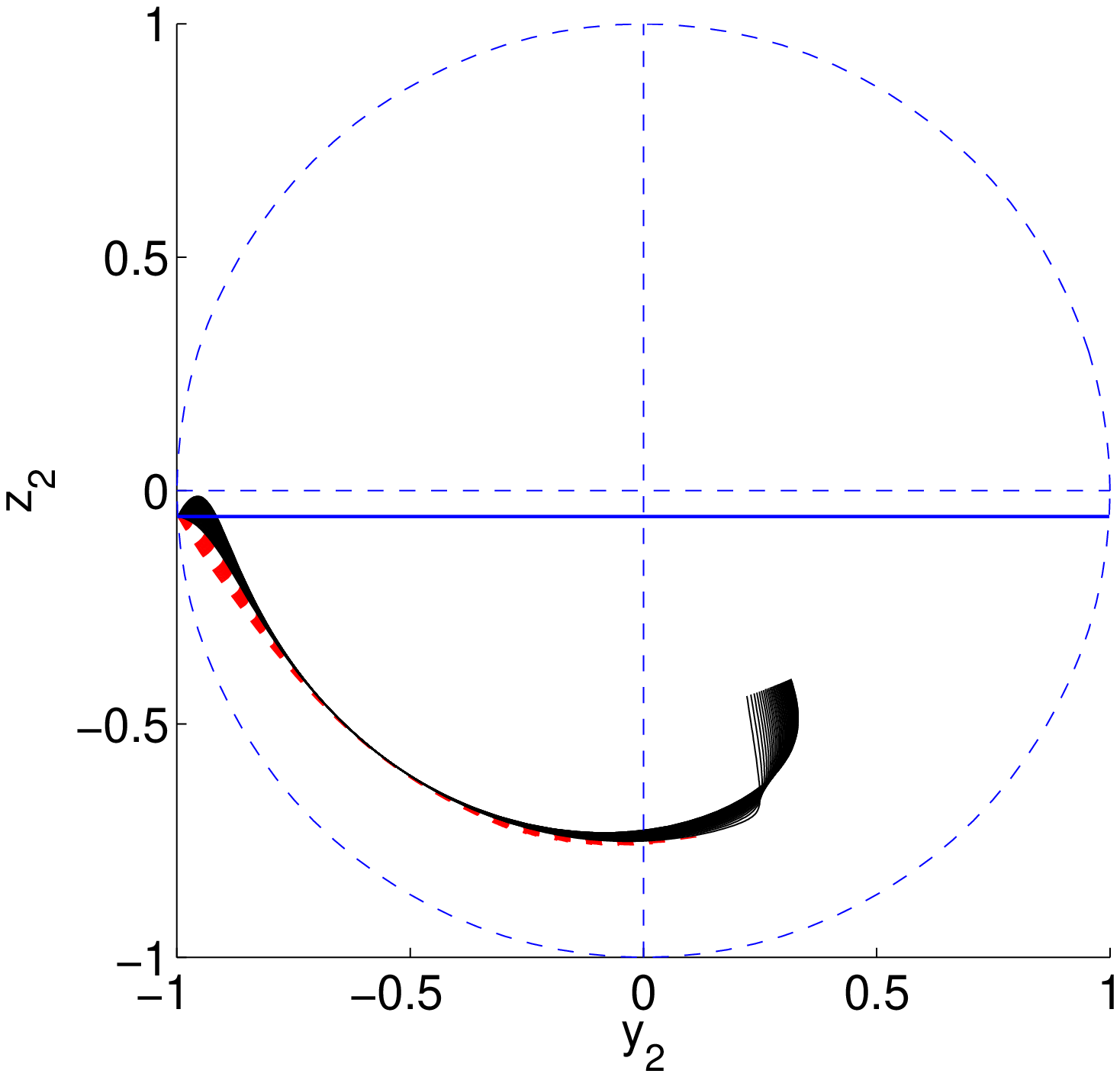}
\caption{\label{fig2} Structure of the projection of the singular
flow onto the planes $(y_1,z_1)$ and $(y_2,z_2)$ in the
cerebrospinal fluid/water case. The trajectories are plotted in
black (solid line) and in red (dashed line). The control fields of
the dashed extremals diverge. The trajectories have been plotted
up to the explosion of the field (The absolute value of the field
is larger than $10^{5}$). The horizontal solid line is a singular
line of the first spin.}
\end{figure}

\begin{figure}
\centering
\includegraphics[width=\sizefig\textwidth]{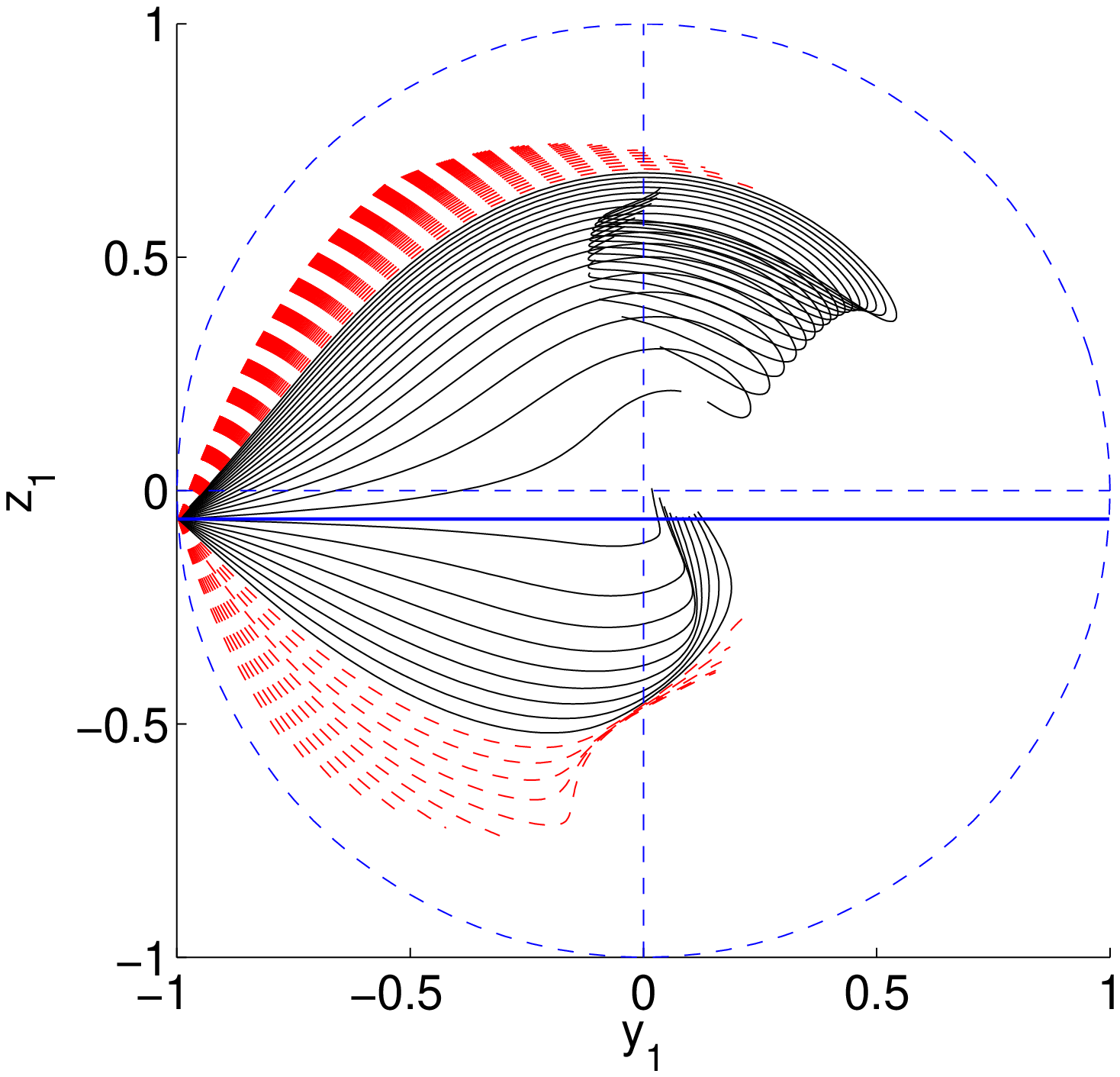}
\includegraphics[width=\sizefig\textwidth]{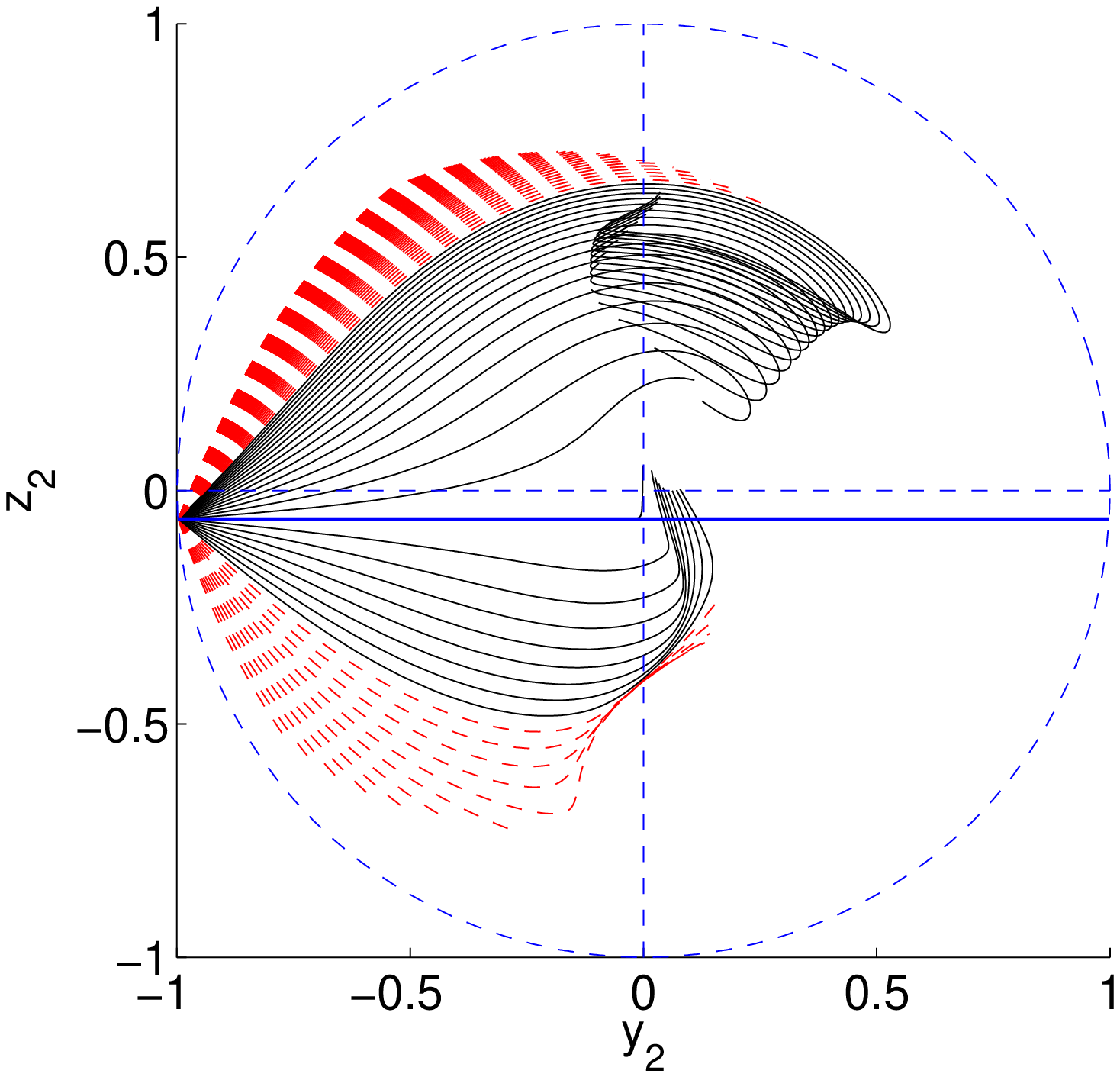}
\caption{\label{fig3} Same as Fig. \ref{fig2} but for the
grey/white matter case. An explosion of the control field is
observed for the red trajectories.}
\end{figure}

\begin{figure}
\centering
\includegraphics[width=\sizefig\textwidth]{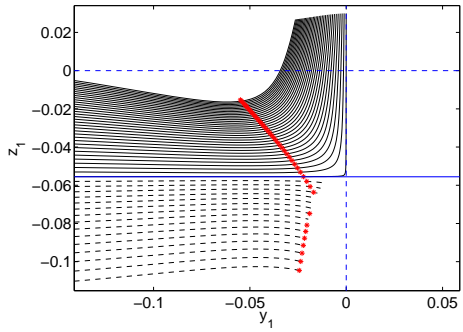}
\caption{\label{fig4} Zoom of the results of Fig. \ref{fig2} for
the spin 1 near the origin. The red crosses indicate the position
of the first conjugate point. The dashed lines represent the
singular trajectories for which the control field diverges.}
\end{figure}

We
present some numerical simulations concerning the singular
trajectories. The projection of $S(q_0)$ on the planes
$(y_1,z_1)$, $(y_2,z_2)$ shows the effect of the relaxation
parameters on the contrast. This point is illustrated by the
figures \ref{fig2} and \ref{fig3} for the cerebrospinal
fluid/water and grey/white matter of cerebrum cases, respectively.
In each example, we assume that a bang pulse of large amplitude
has been first applied to the system, the initial point of the
singular flow is of coordinates
$((-\sqrt{1-z_0^2},z_0),(-\sqrt{1-z_0^2},z_0))$ where $z=z_0$ is
the horizontal singular line of the first spin. This first bang is
necessary so that the singular trajectory of the spin 1 can reach
the center of the Bloch ball. One clearly sees in Fig. \ref{fig3}
the similar structure of the different singular trajectories of
the two spins. The situation is completely different in Fig.
\ref{fig2} for the first example. This explains the excellent and
weak contrasts that can be reached in the first and second
examples with an optimal sequence of the form bang-singular. Note
that some singular control fields diverge as displayed in Figs.
\ref{fig2} and \ref{fig3}. The conjugate points defined in Sec.
\ref{sec2} have been computed for each singular extremal as shown
in Fig. \ref{fig4}. Similar results have been obtained for the
spin 2 and for the grey/white matter case. This shows that the
structure bang-singular is not optimal since the first conjugate
point occurs before the saturation of the spin. A more complicated
pulse sequence such as bang-singular-bang-singular has therefore
to be used.
\subsection{Some preliminaries numerical results on the contrast problem}
Due to the numerical difficulty of the computation of the
bang-singular-bang-singular optimal sequence, we only present
in this section some preliminary results.
To make the numerical simulations, we have used a differential
continuation method of the Hampath code \cite{hampath} where the
cost is regularized by adding a L$^2$ (or a L$^{2-\lambda}$)
penalty on the control. Note that another continuation on the
transfer time has also been used in the computations. Such results
can be compared with the GRAPE algorithm \cite{gershenzon,khaneja2,skinner,skinner2} which is a
standard approach in NMR to solve the optimization problems.

In the first study, the cost is regularized as
$$
C(x(T))+(1-\lambda) \int_0^Tu^2(t)dt,
$$
where $\lambda$ is a continuation parameter and the transfer time
varies starting from $T_{min}+\varepsilon$ to $2T_{min}$, where
$T_{min}$ is the minimum time to saturate the first spin and
$\varepsilon\ll 1$ an arbitrary parameter. According to Sec.
\ref{sec3}, in the limit case where $T=T_{min}$, the optimal
solution of the contrast problem is exactly the solution of
driving the first spin to the origin.

\begin{figure}
\centering
\includegraphics[width=\sizefig\textwidth]{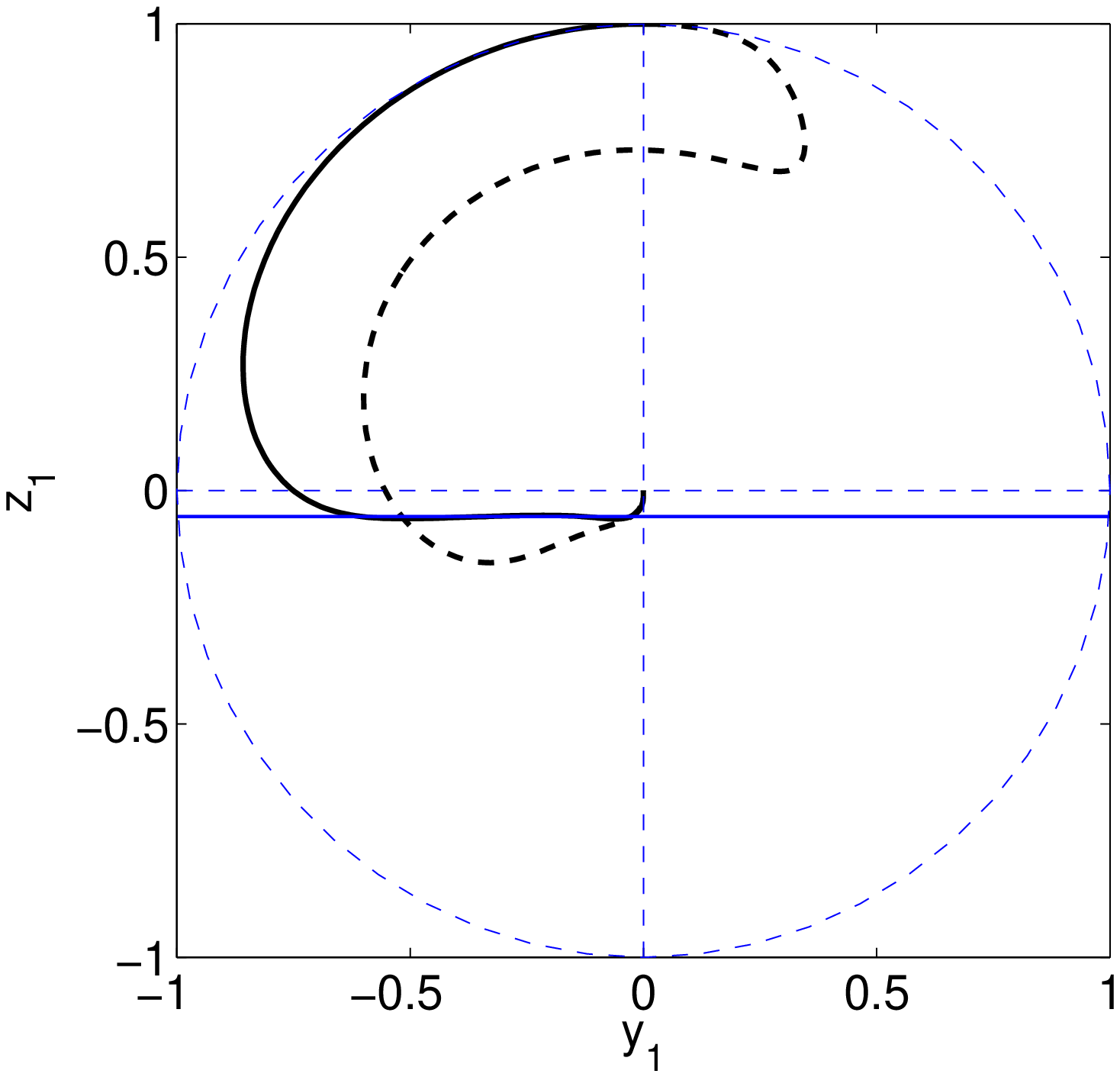}
\includegraphics[width=\sizefig\textwidth]{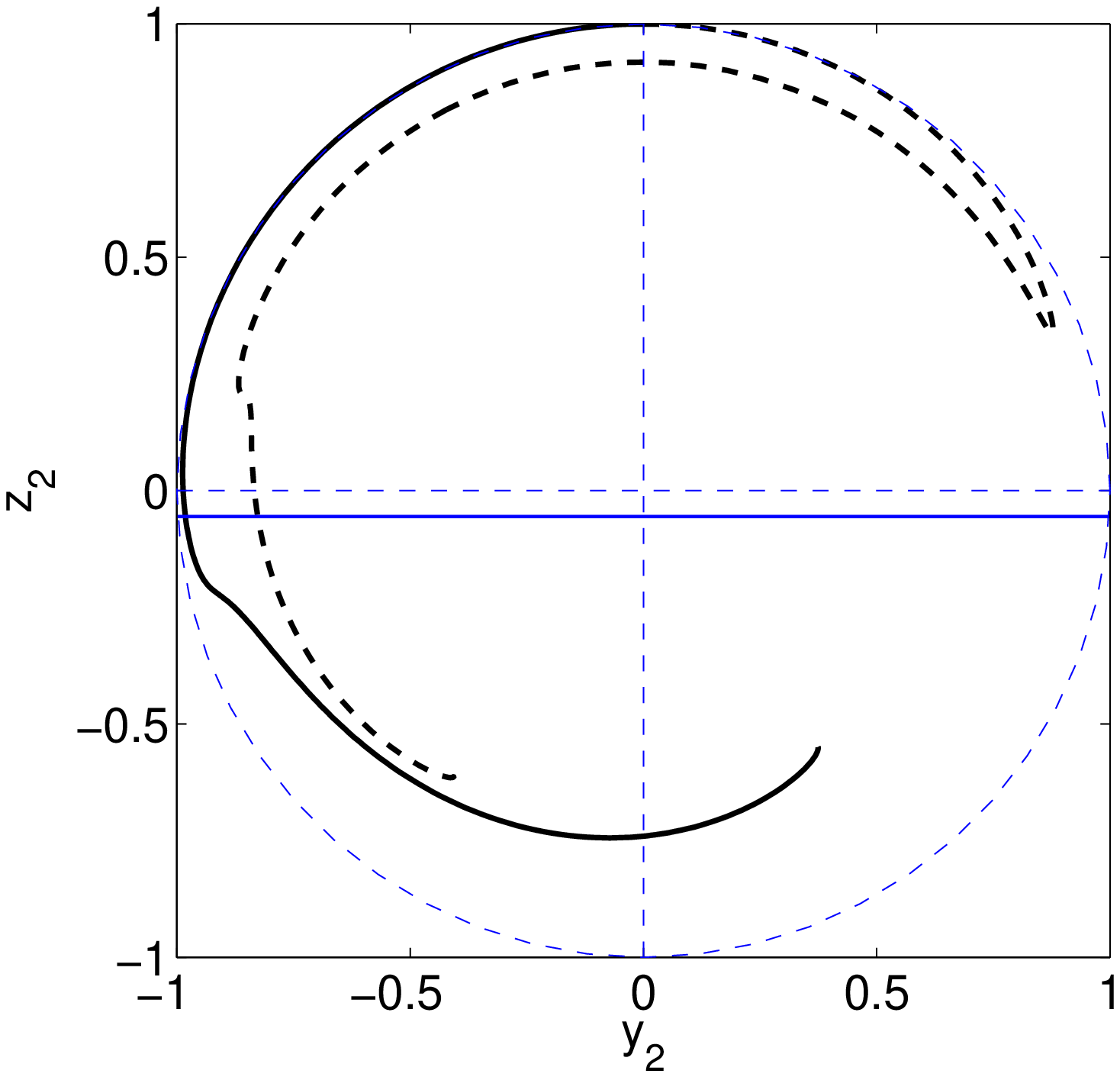}
\caption{\label{fig5} (The cerebrospinal fluid/water case) Trajectories of the first and second spins
for $T_{min}+\varepsilon$ and $2T_{min}$ in solid and dashed
lines, respectively. The horizontal singular line is plotted in
solid line. The parameter $\lambda$ is taken as 0.9.}
\end{figure}

\begin{figure}
\centering
\includegraphics[width=\sizefig\textwidth]{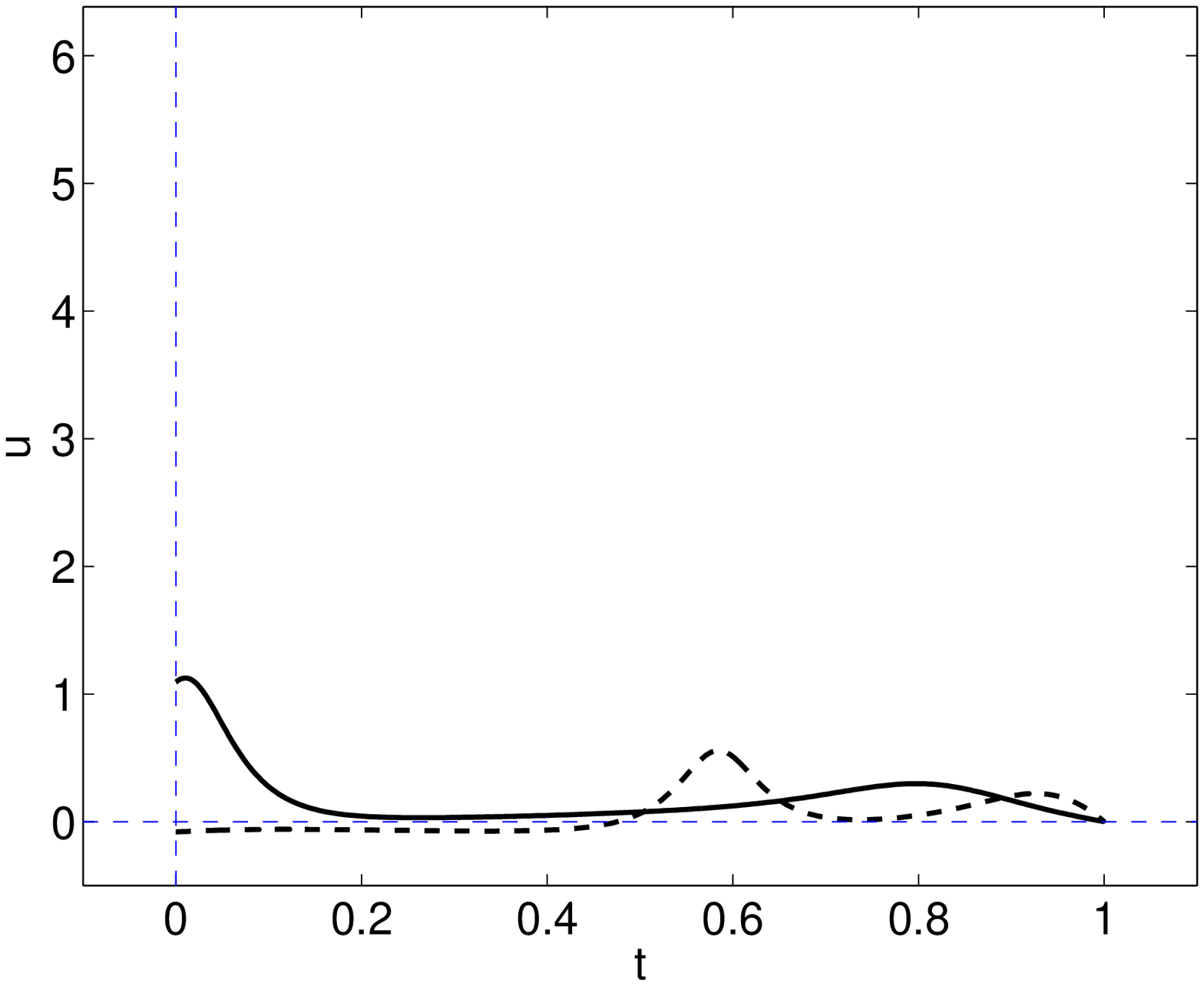}
\includegraphics[width=\sizefig\textwidth]{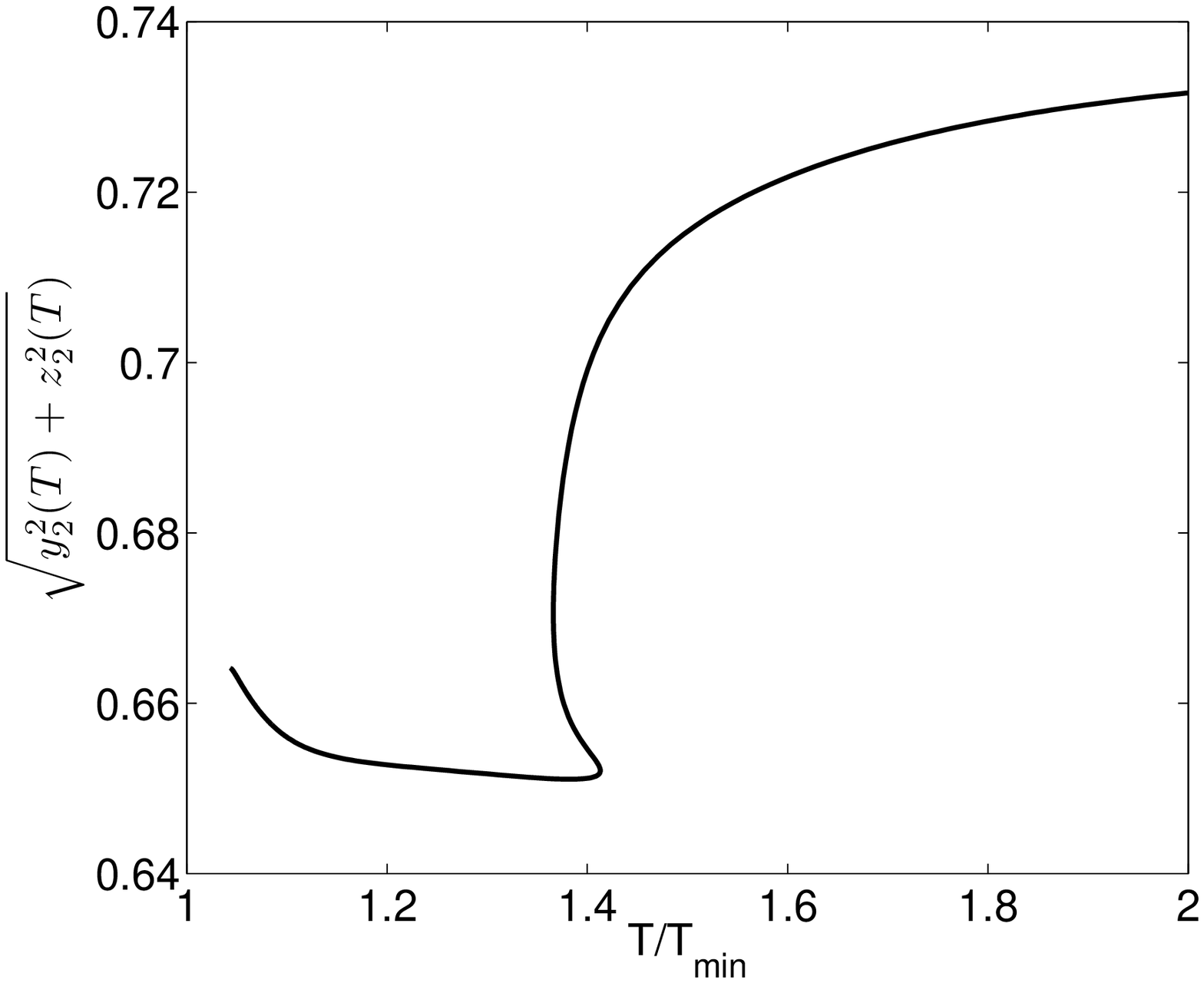}
\caption{\label{fig6} (top) Evolution of the control field
associated to Fig. \ref{fig5} for $T_{min}+\varepsilon$ and
$2T_{min}$ in solid and dashed lines, respectively. The time $T$
has been normalized to 1 to plot the two control fields on the
same figure. (bottom) Evolution of the contrast parameter
$\sqrt{y_2(T)^2+z_2(T)^2}$ as a function of the control duration.}
\end{figure}

\begin{figure}
\centering
\includegraphics[width=\sizefig\textwidth]{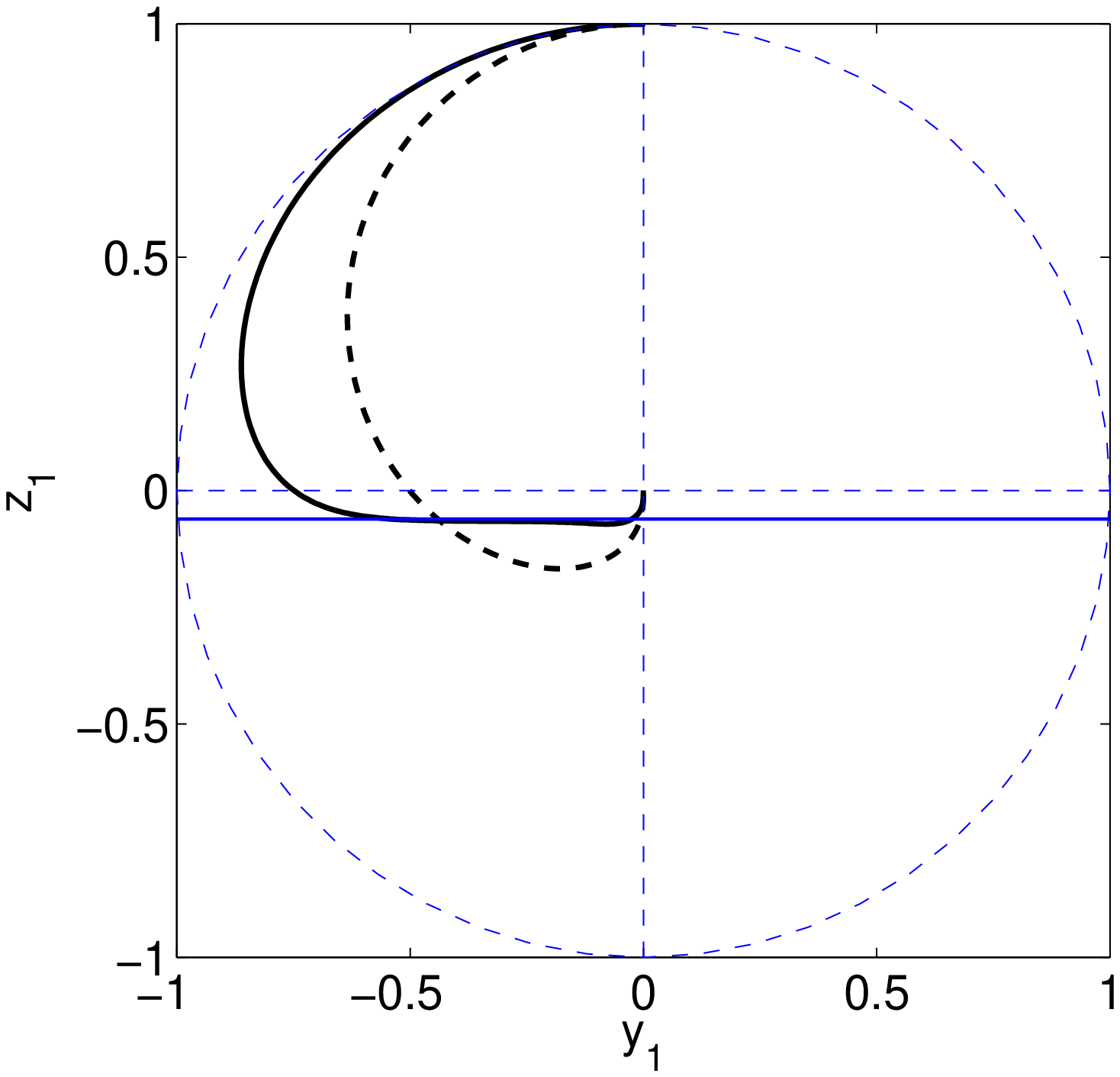}
\includegraphics[width=\sizefig\textwidth]{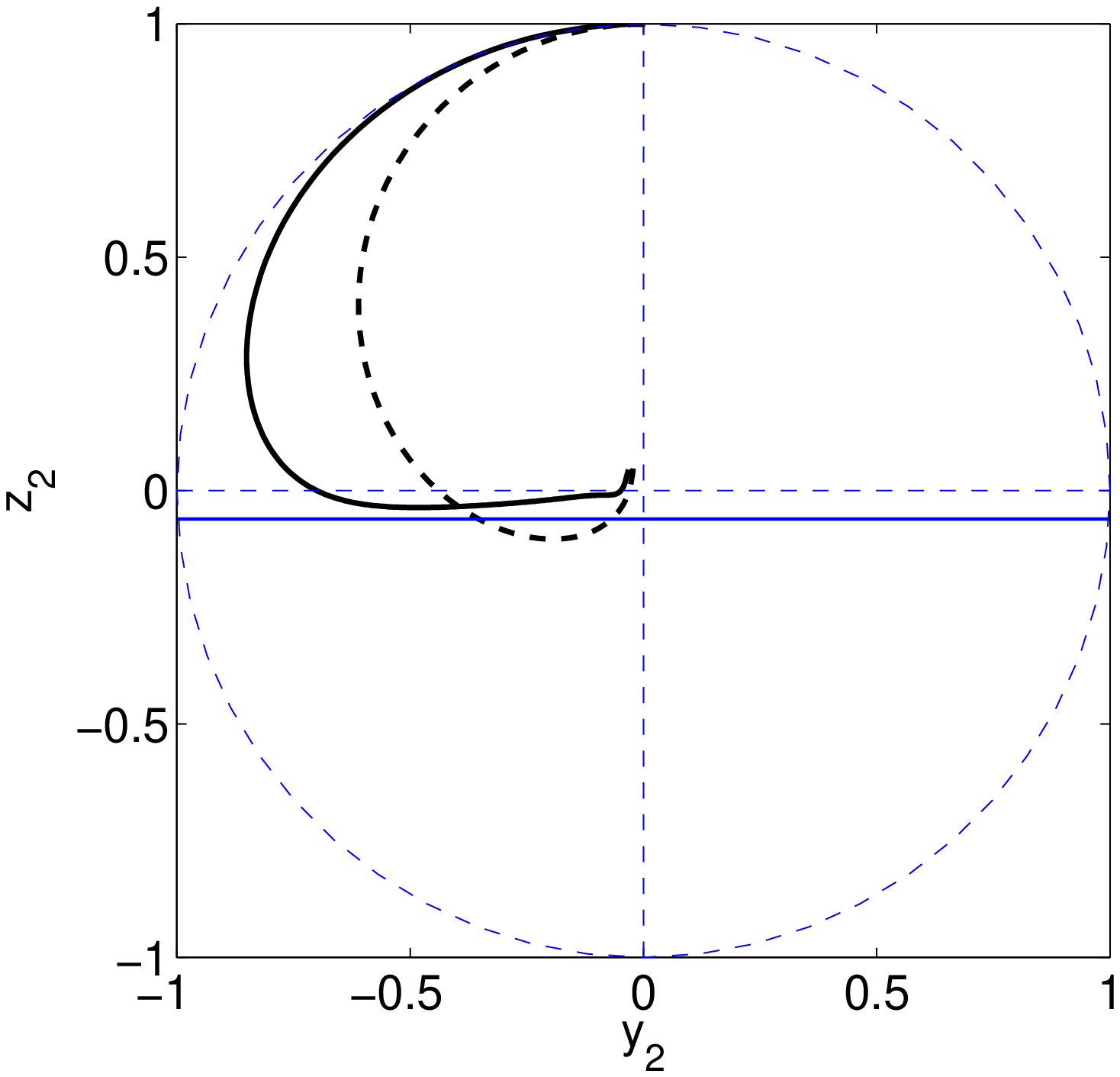}
\caption{\label{fig7} Same as Fig. \ref{fig5} but for the
grey/white matter of cerebrum.}
\end{figure}

\begin{figure}
\centering
\includegraphics[width=\sizefig\textwidth]{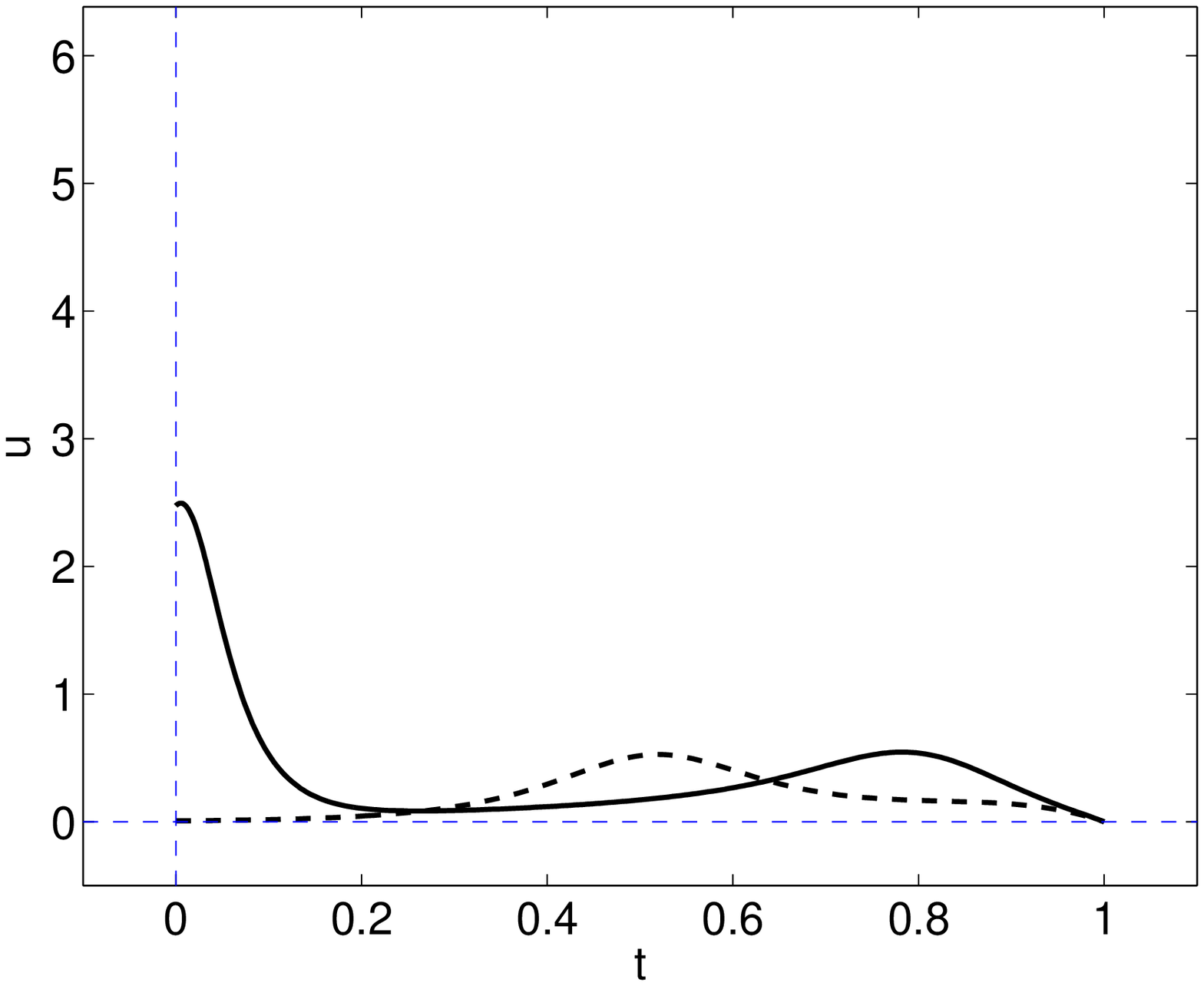}
\includegraphics[width=\sizefig\textwidth]{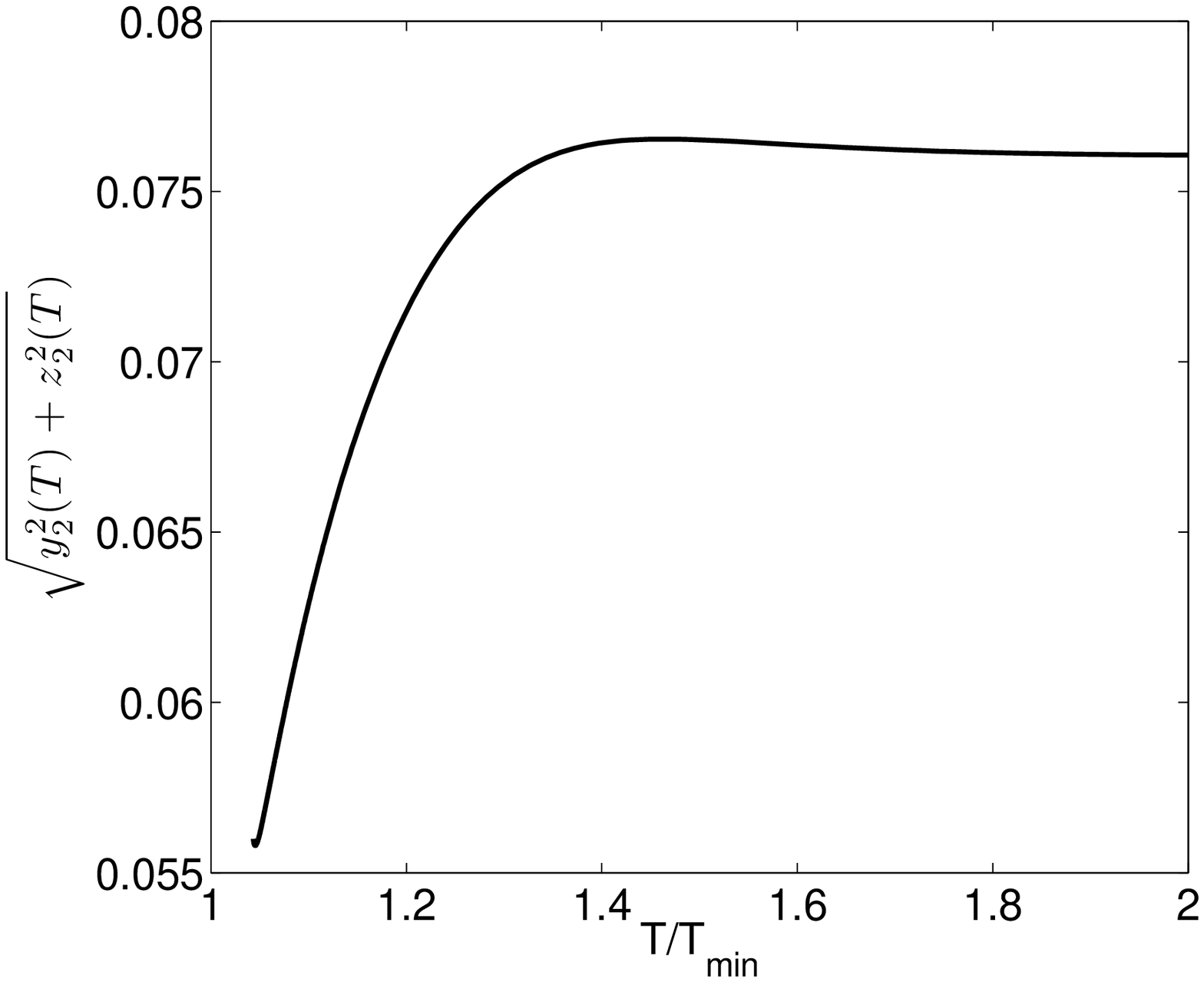}
\caption{\label{fig8} Same as Fig. \ref{fig6} but for the
grey/white matter of cerebrum.}
\end{figure}

The different numerical results are presented in Figs. \ref{fig5},
\ref{fig6}, \ref{fig7} and \ref{fig8}. The different behaviors for
the two examples can be clearly seen since the best contrast
$\sqrt{y_2^2(t)+z_2^2(t)}$ is of the order of 0.73 and 0.07 in the
first and second examples, respectively. Note also that for
$T=T_{min}+\varepsilon$, the trajectory of the spin 1 is very
close to the trajectory for saturating this spin in minimum time.

An interesting phenomenon can be observed in the cerebrospinal
fluid/water situation in Fig. \ref{fig5}, where there exists a
bifurcation of the optimal policy when $T$ increases. This is
related to the introduction of a bang-bang policy associated to a
SiCo singularity. Also we observe that the optimal policy is
crossing the $z_1$- axis of revolution. Further work is necessary
to improve the continuation method near $\lambda=1$ since the
L$^2$- regularization of the cost is not adapted to the control
saturation that can be found in the SiSi singularity where the
control is L$^1$ and not L$^2$.


This point is illustrated by a second series of simulations where
we have considered the following regularized cost:
$$
C(x(T))+(1-\lambda)\int_0^T|u|^{2-\lambda}(t)dt
$$
and as before a continuation has been performed on the control
duration $T$. The computation has been done for $\lambda=0.9$.
Similar contrasts have been reached in this second situation.
Note, however, the different peaks appearing in the evolution of
$u$, to be compared to the first regularization.

We complete this paper by illustrating our numerical results on a
simulated and a real contrast experiments. For the simulated experiment, we consider two surfaces as
displayed in Fig. \ref{fig11} filled in with spins 1 or 2 in a
homogeneous manner. We apply the optimal control field and we
associate a color to the final modulus of the magnetization vector
of the spin 2. This color is white if the modulus is equal to 1,
black if it is zero and a grey variant between. One clearly sees
in Fig. \ref{fig11} the excellent and weak contrasts that can be
obtained in the first and second examples.

\begin{figure}
\centering
\includegraphics[width=\sizefig\textwidth]{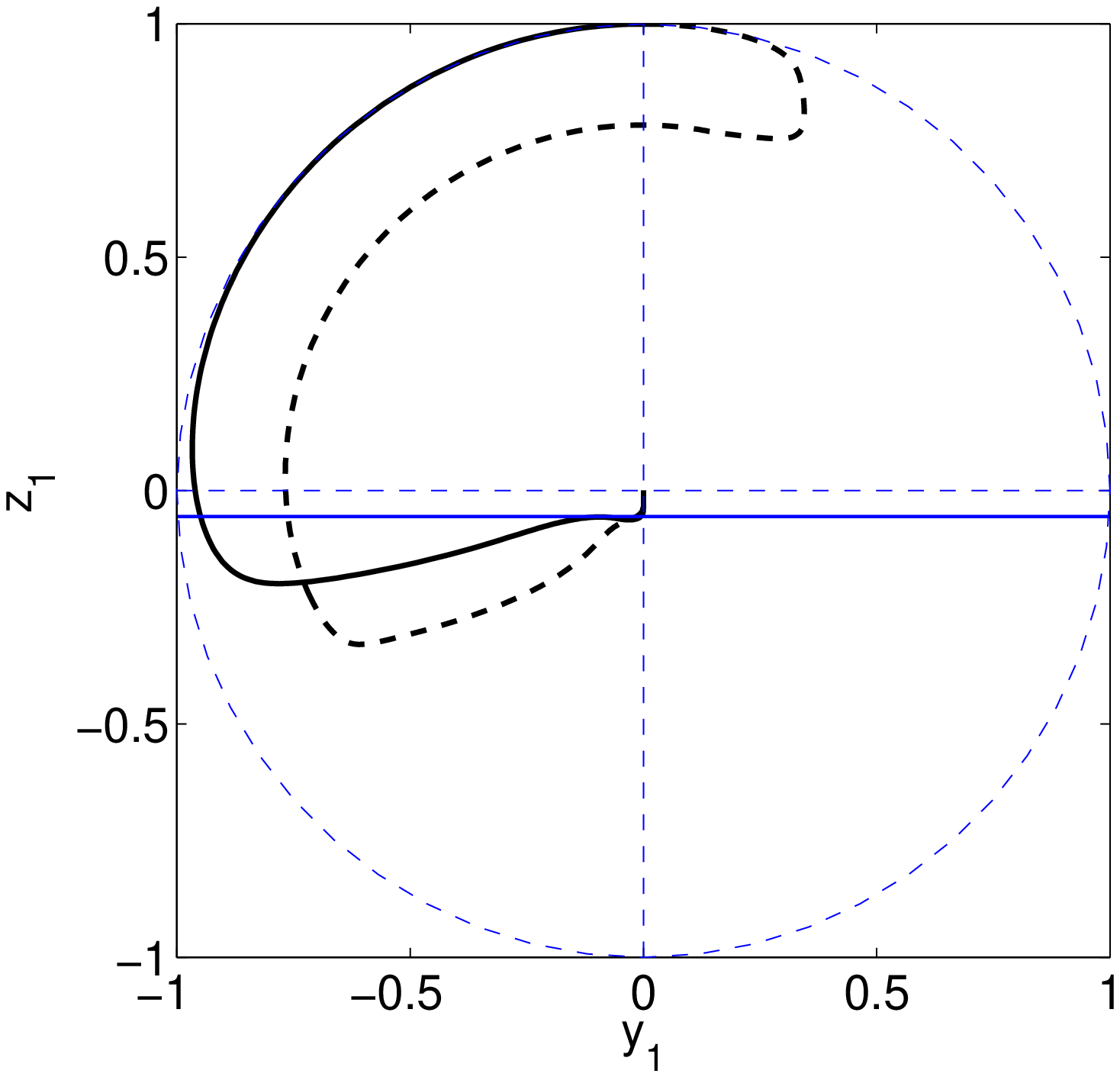}
\includegraphics[width=\sizefig\textwidth]{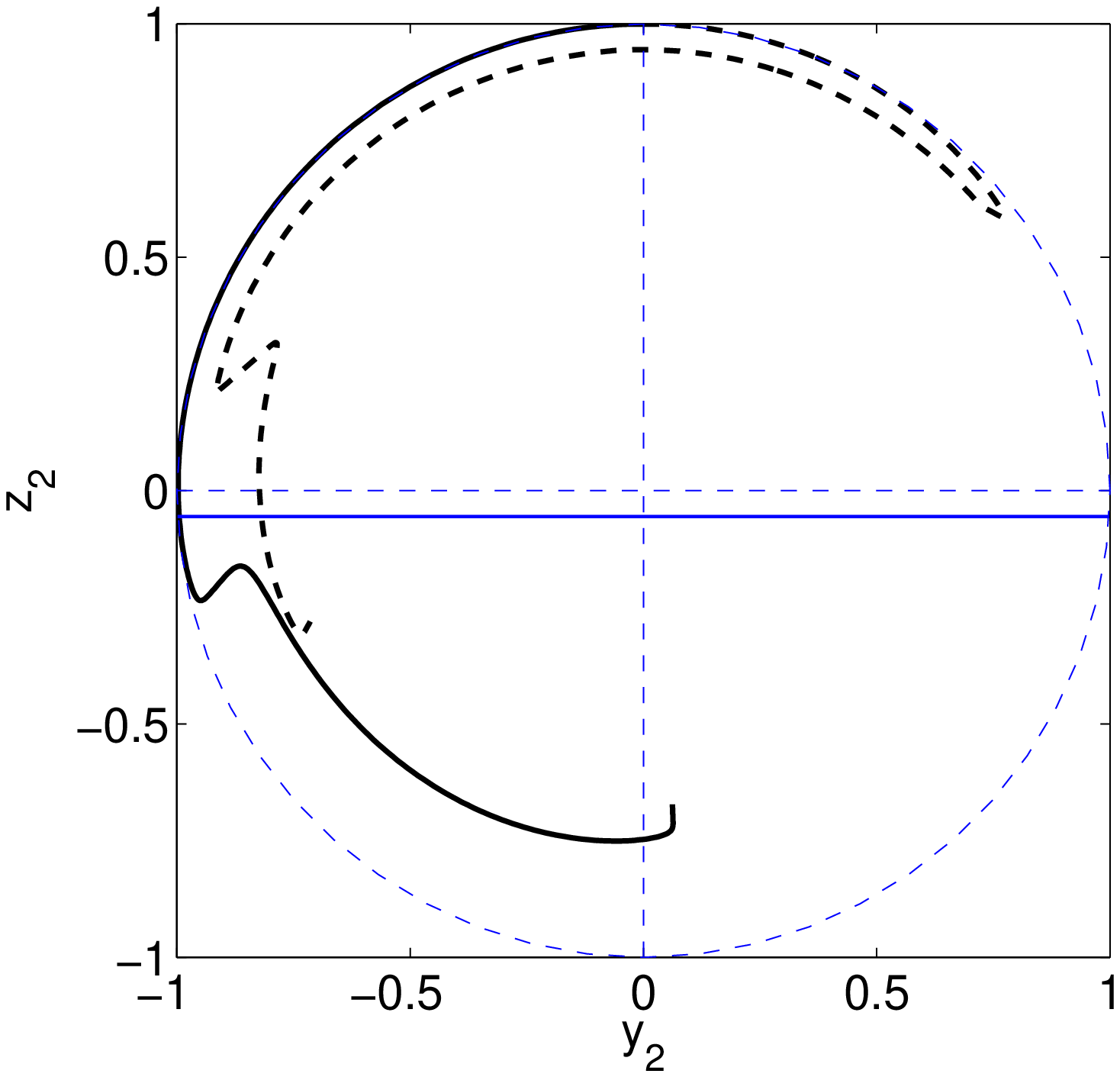}
\caption{\label{fig9} Same as Fig. \ref{fig5} but for the second
regularized cost functional. The parameter $\lambda$ is taken as
0.93.}
\end{figure}
\begin{figure}
\centering
\includegraphics[width=\sizefig\textwidth]{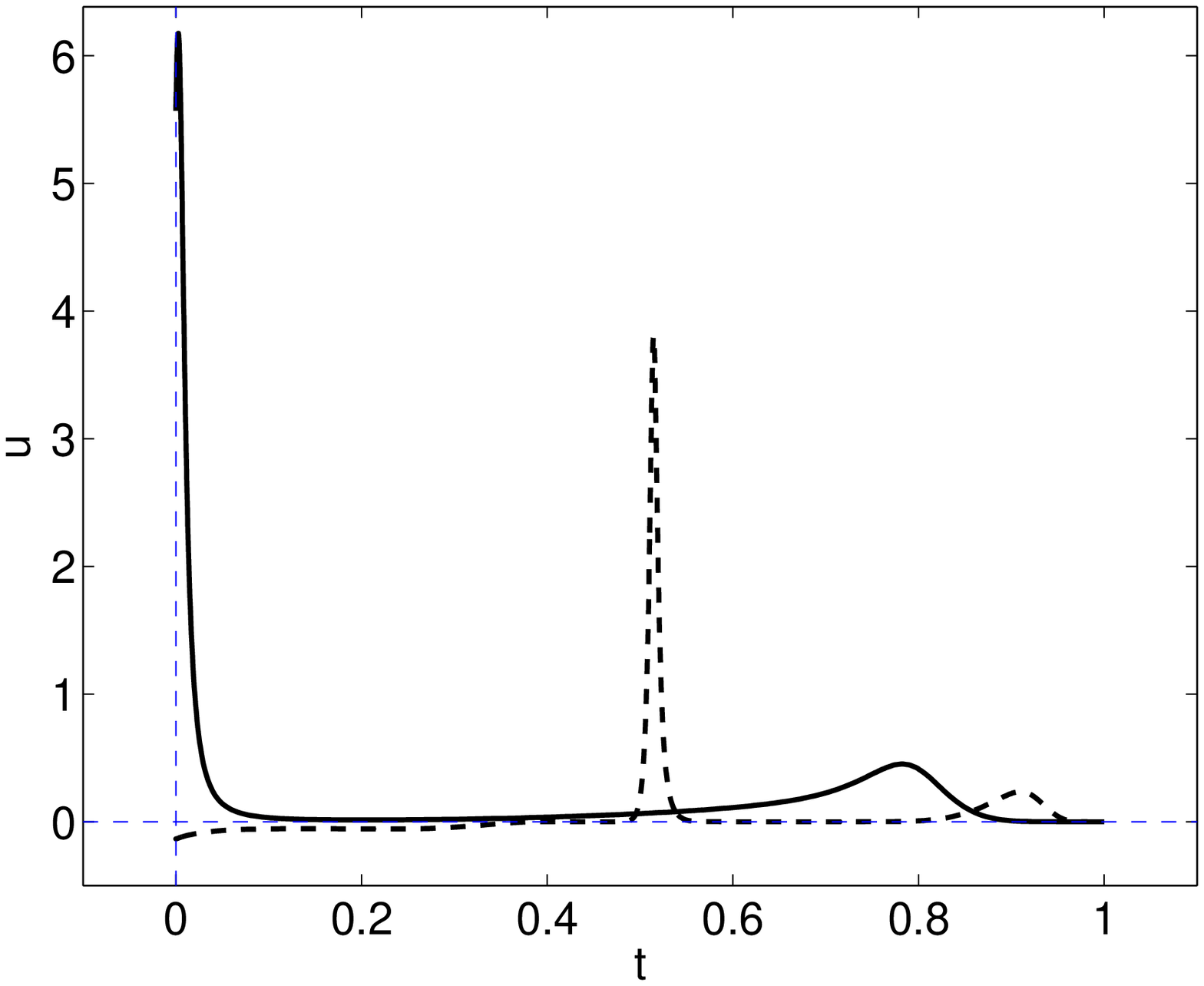}
\includegraphics[width=\sizefig\textwidth]{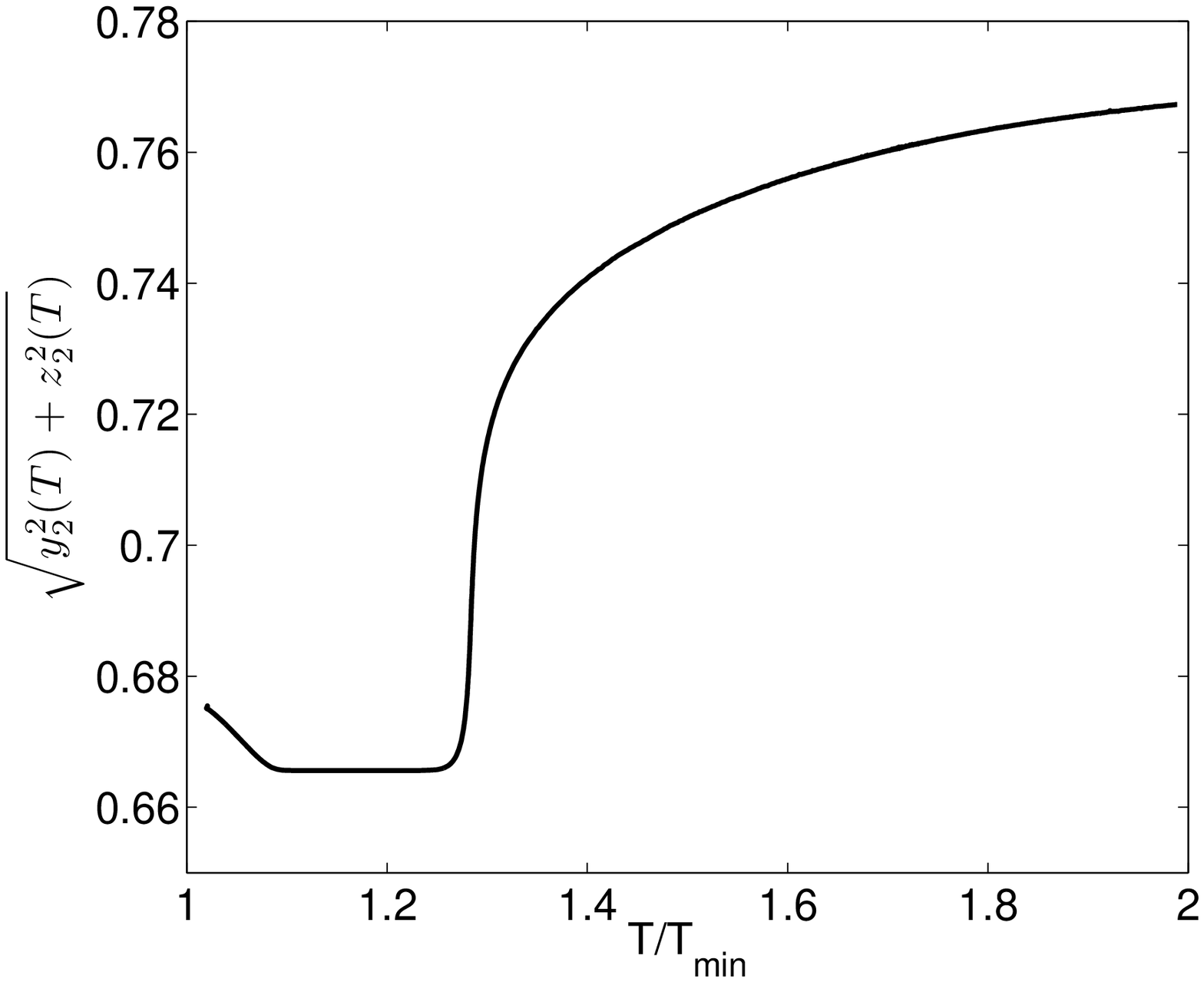}
\caption{\label{fig10} Same as Fig. \ref{fig6} but for the second
regularized cost functional.}
\end{figure}

\begin{figure}
\centering
\includegraphics[width=\sizefig\textwidth]{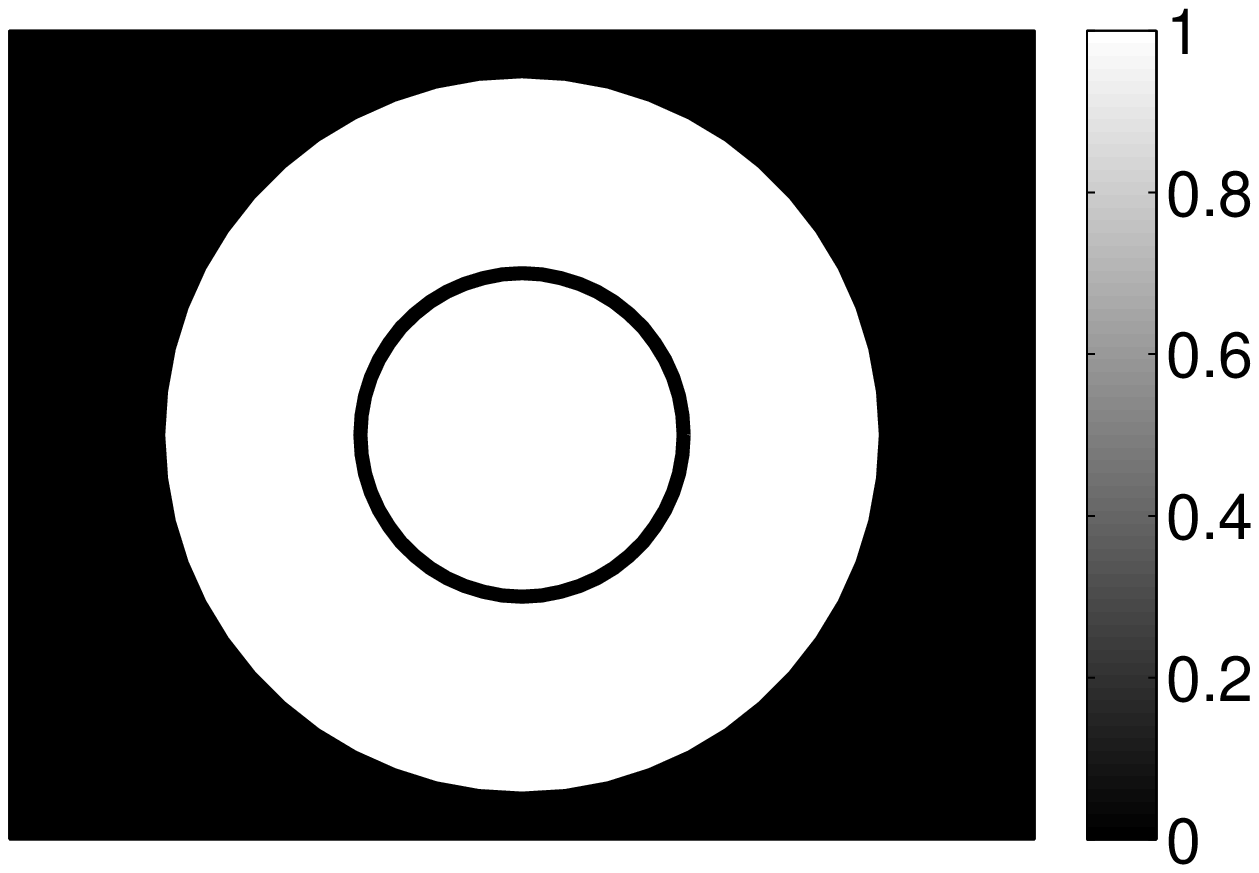}
\includegraphics[width=\sizefig\textwidth]{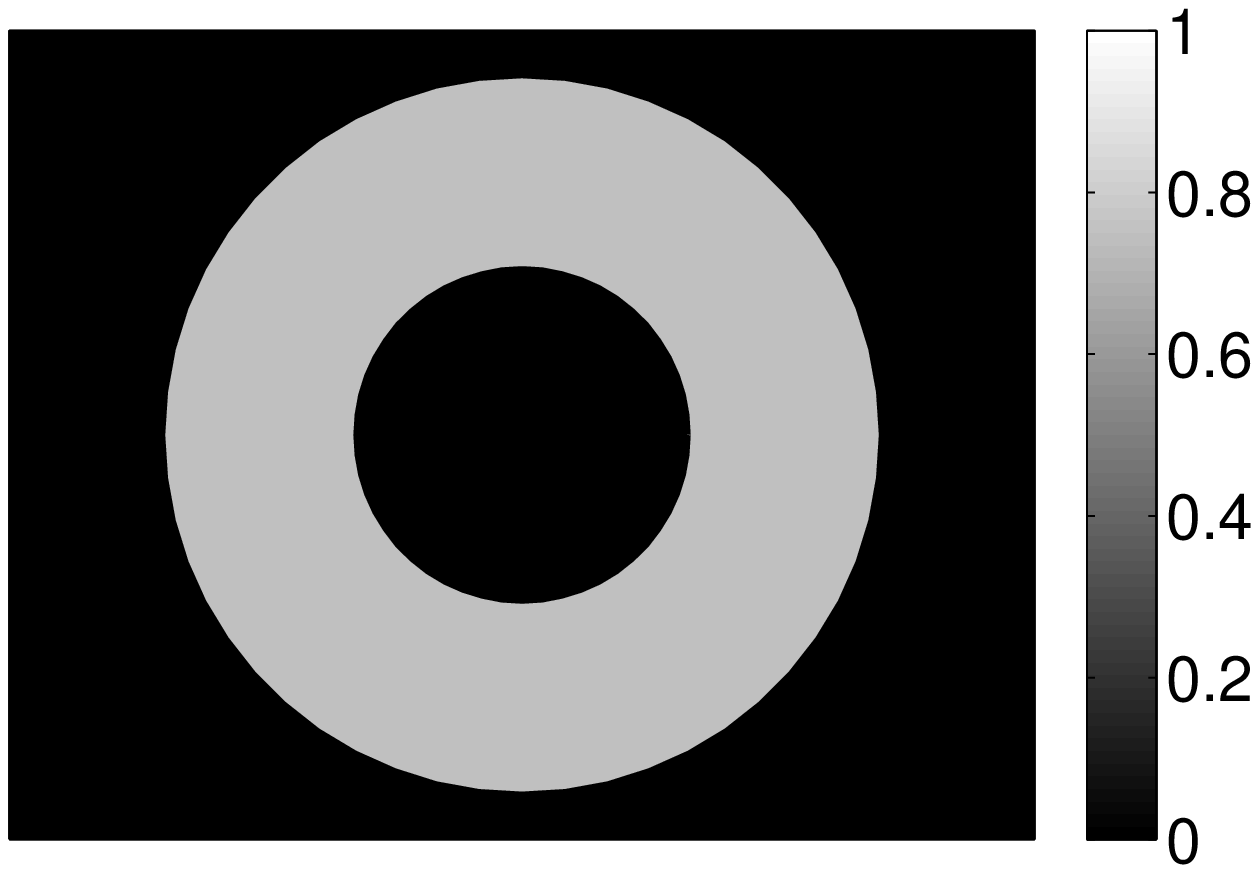}
\includegraphics[width=\sizefig\textwidth]{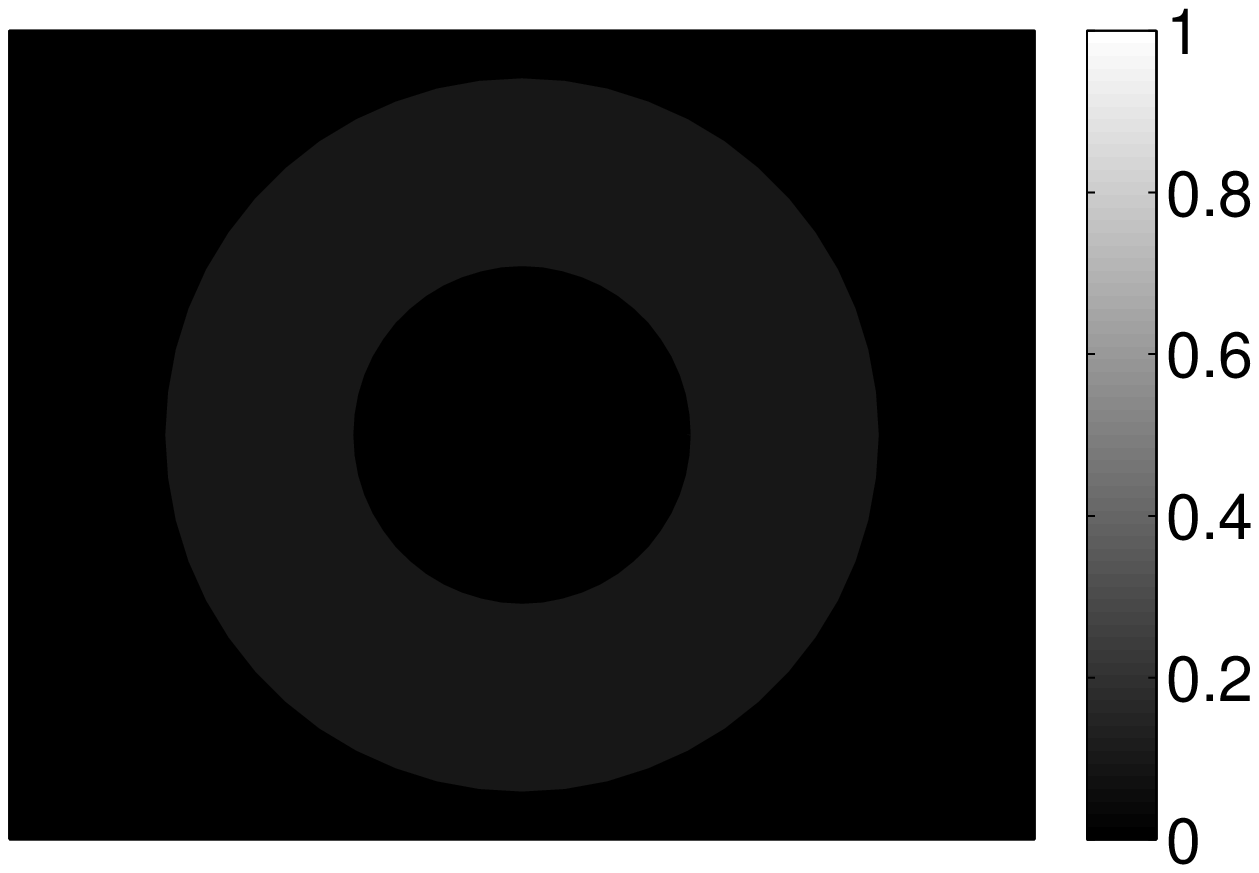}
\caption{\label{fig11} Simulated experimental results on the
contrast problems of the cerebrospinal fluid/water (middle) and
the grey/white matter of cerebrum (bottom) examples. The inner
disk mimics the spin 1, while the outside ring mimics the spin 2.
The two surfaces are separated by a thin black circle. The top
figure is a reference image where a 90 degree pulse has been applied to the two spins. The middle and bottom images are a representation of the
contrast as could be done in a real experiment. For these two images, the control sequence is the optimal field. A color has been
associated to each value of the contrast between 0 and 1, 0 and 1
corresponding respectively to the colors black and white.}
\end{figure}

\begin{figure}
\centering
\includegraphics[width=\sizefig\textwidth]{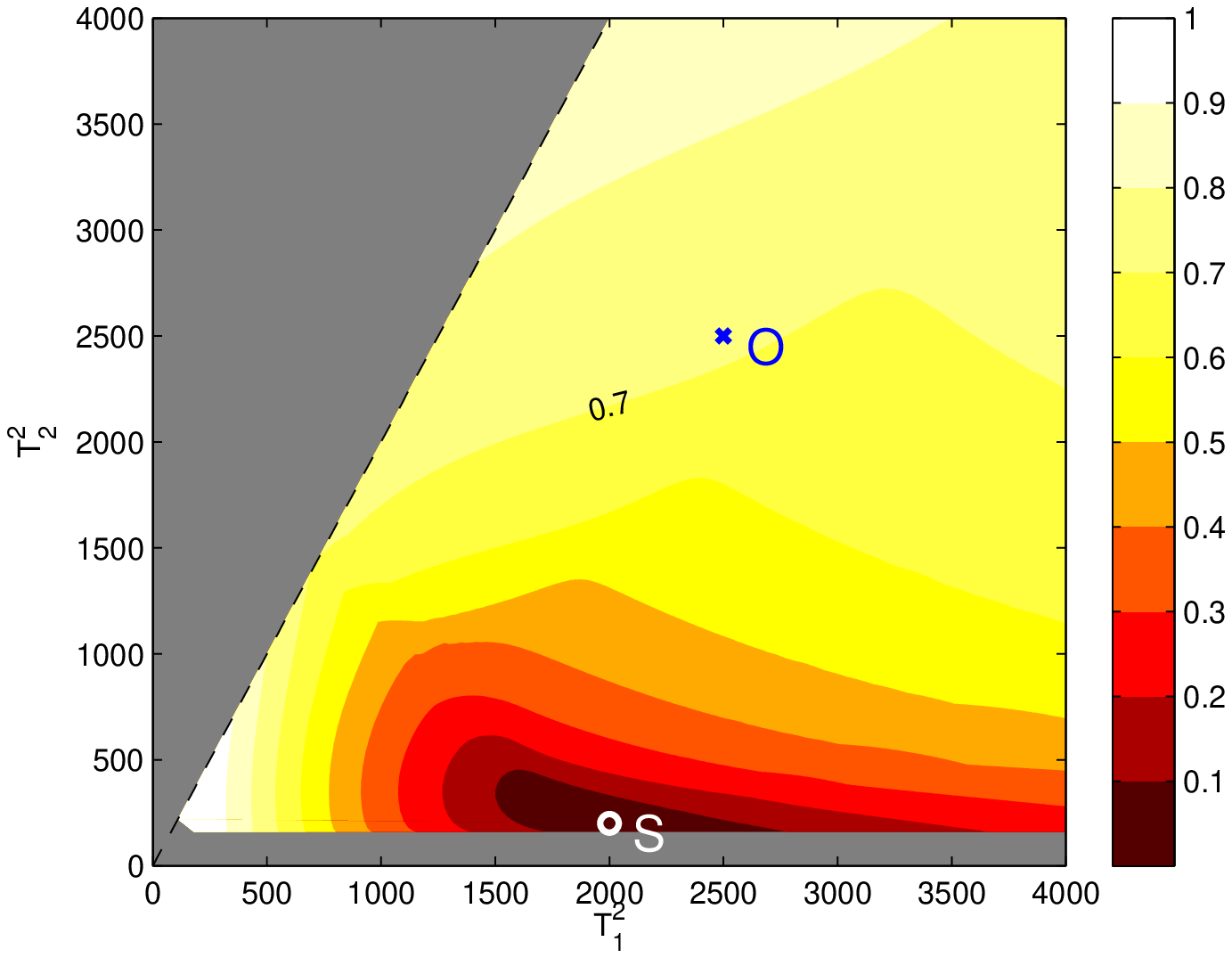}
\includegraphics[width=\sizefig\textwidth]{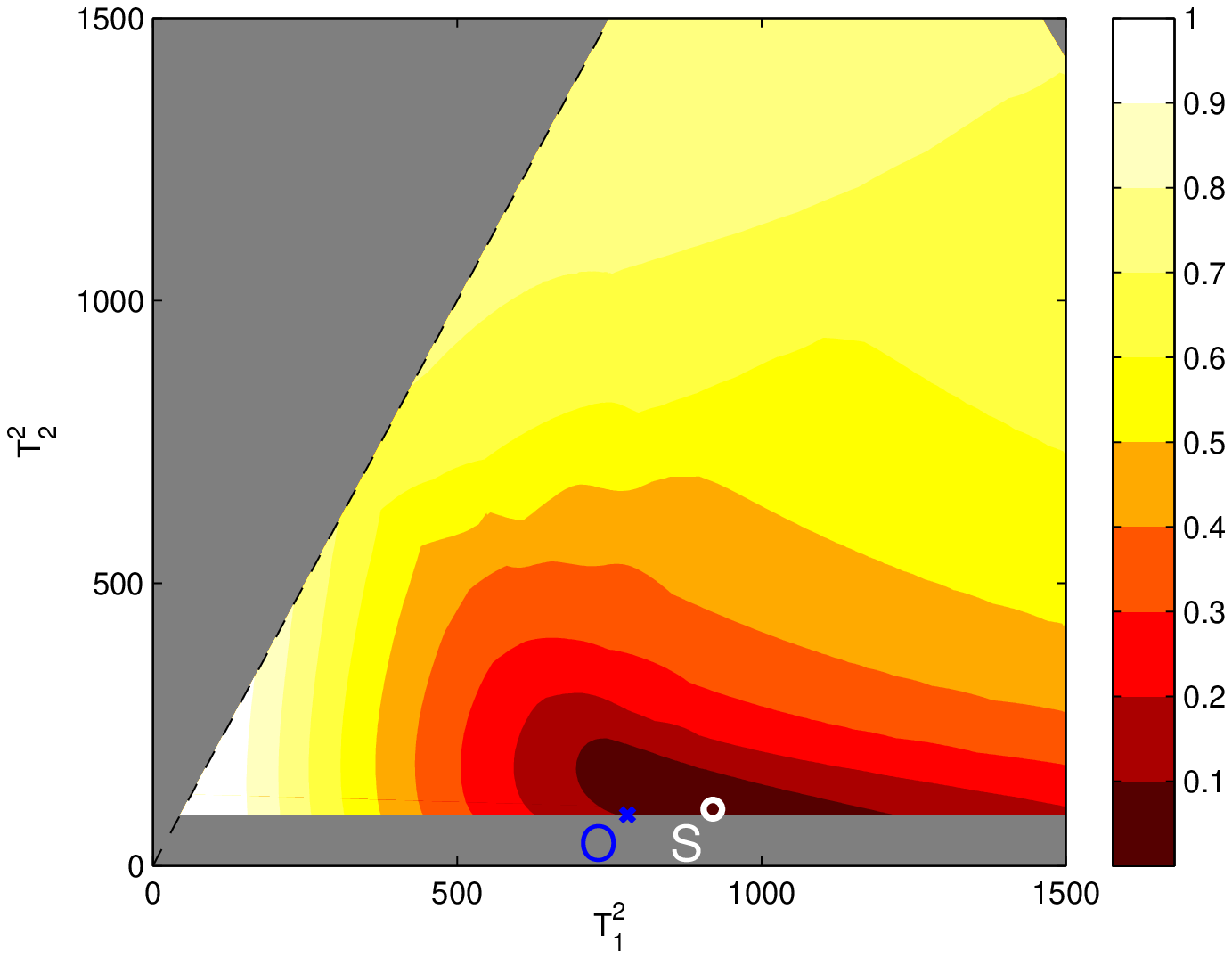}
\includegraphics[width=\sizefig\textwidth]{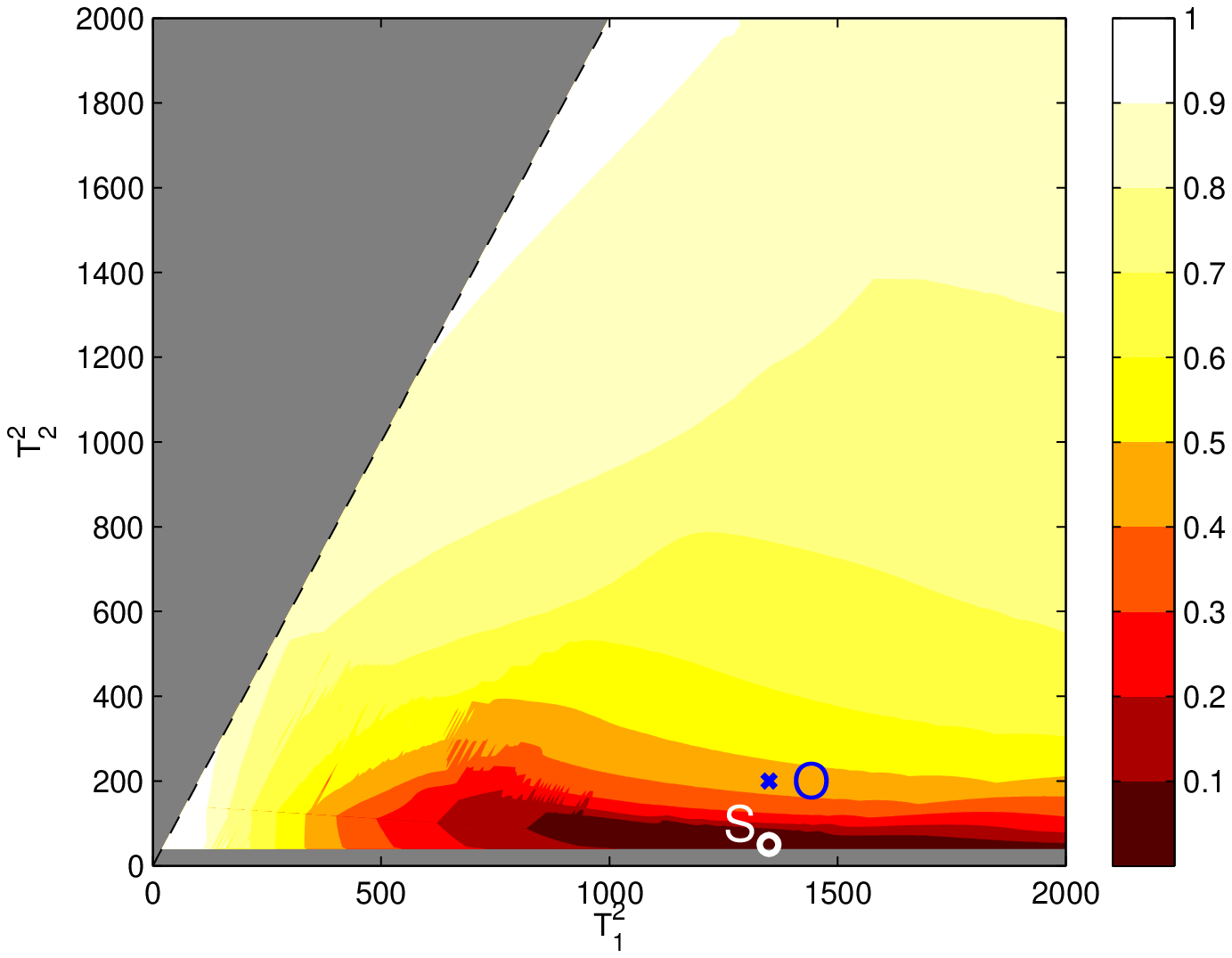}
\caption{\label{fig12}  Interpolation results for the (top) Fluid case, the (middle) Gray matter case and the (bottom) Blood case.
The contrast is computed with respect to the spin 2 relaxation parameters. The parameters of the spin 1 are fixed to the ones of the points $S$. The points $O$ give us the contrast
for known problems, as the fluid/water case on the top figure, gray/white matter case on
the middle one and deoxygenated/oxygenated blood on the bottom figure.}
\end{figure}

In Fig. \ref{fig12}, we compute by interpolation the contrast between
the two spin particles for different values of the relaxation parameters. To have a 2-D representation, one fixes the first spin
parameters. In the top figure, the first spin corresponds to the cerebrospinal fluid
($T_1^1=2000$ ms and $T_2^1=200$ms), in the middle figure,
it is the gray matter ($T_1^1=920$ ms and $T_2^1=100$ ms)
and in the bottom one, it is the deoxygenated blood ($T_1^1=1350$ ms and $T_2^1=50$ ms), this latter example being illustrated below experimentally.
In each case, we fix the control duration $T$ to $1.5T_{min}$ and we choose
the regularized cost $C(x(T))+(1-\lambda) \int_0^T|u|^{2-\lambda}(t)dt,$
with $\lambda=0.9$. We consider the following variations for the parameters of the
second spin:
\begin{eqnarray*}
& & x_{min} \leq T_1^2 \leq x_{max} \\
& & y_{min} \leq T_2^2 \leq y_{max},
\end{eqnarray*}
where $(x_{min},x_{max},y_{min},y_{max}) = (80,4000,160,4000)$ for the
fluid case, $(45,1500,90,1500)$ for the matter case and $(20,2000,40,2000)$
for the blood case. The linear inequalities $T_2\leq 2T_1$ due to the physical model, leads to convex polyhedrons in Fig. \ref{fig12}.
The starting point of the forthcoming homotopies corresponds to $S=(T_1^1,T_2^1)$
for which the contrast is zero. Then we discretize the edges
of the polytope P into $n$ points and for each $F_k=(T_1^{2,k},T_2^{2,k})$,
$k=1,...,n$, we perform a linear homotopy from $S$ to $F_k$ by introducing a
parameter $\lambda$ such that :
\begin{eqnarray*}
& & T_1^2 = T_1^1 + \lambda (T_1^{2,k} - T_1^1)\\
& & T_2^2 = T_2^1 + \lambda (T_2^{2,k} - T_2^1).
\end{eqnarray*}
At the end, we have $n$ lines starting from $S$ which mesh $P$ and to complete the
figures, we use a standard Matlab interpolation function. We can see in Fig. \ref{fig12},
on the top figure, that the contrast in $O=(2500,2500)$, which corresponds
to the Fluid/Water case, is nearly $0.7$ and on the middle one, in $O=(780,90)$,
the Gray/White matter case, the contrast is almost $0.1$, which agree with the results
given in Figs. \ref{fig10} and \ref{fig8} respectively.

\begin{figure}
\centering
\includegraphics[width=\sizefig\textwidth]{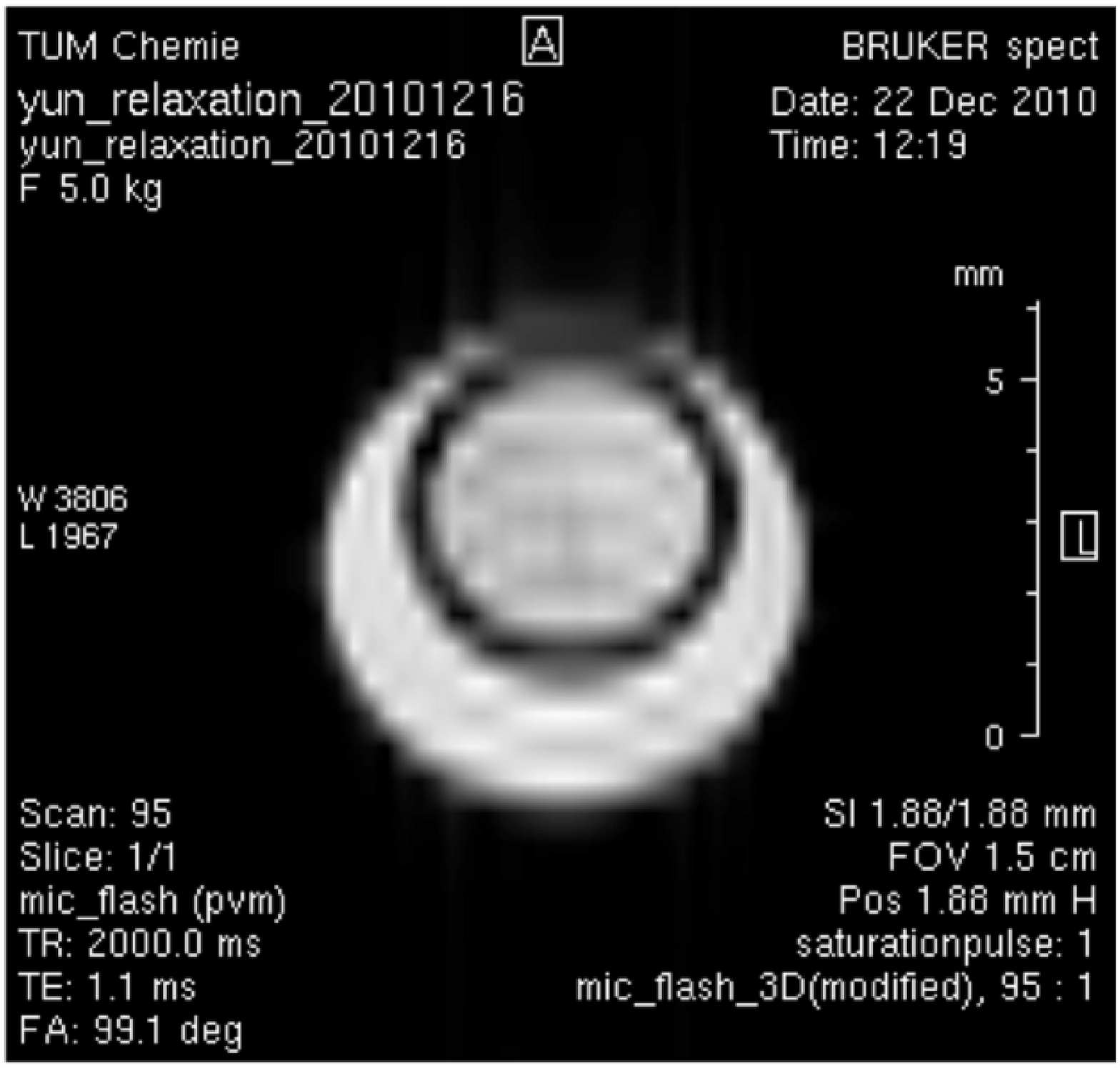}
\includegraphics[width=\sizefig\textwidth]{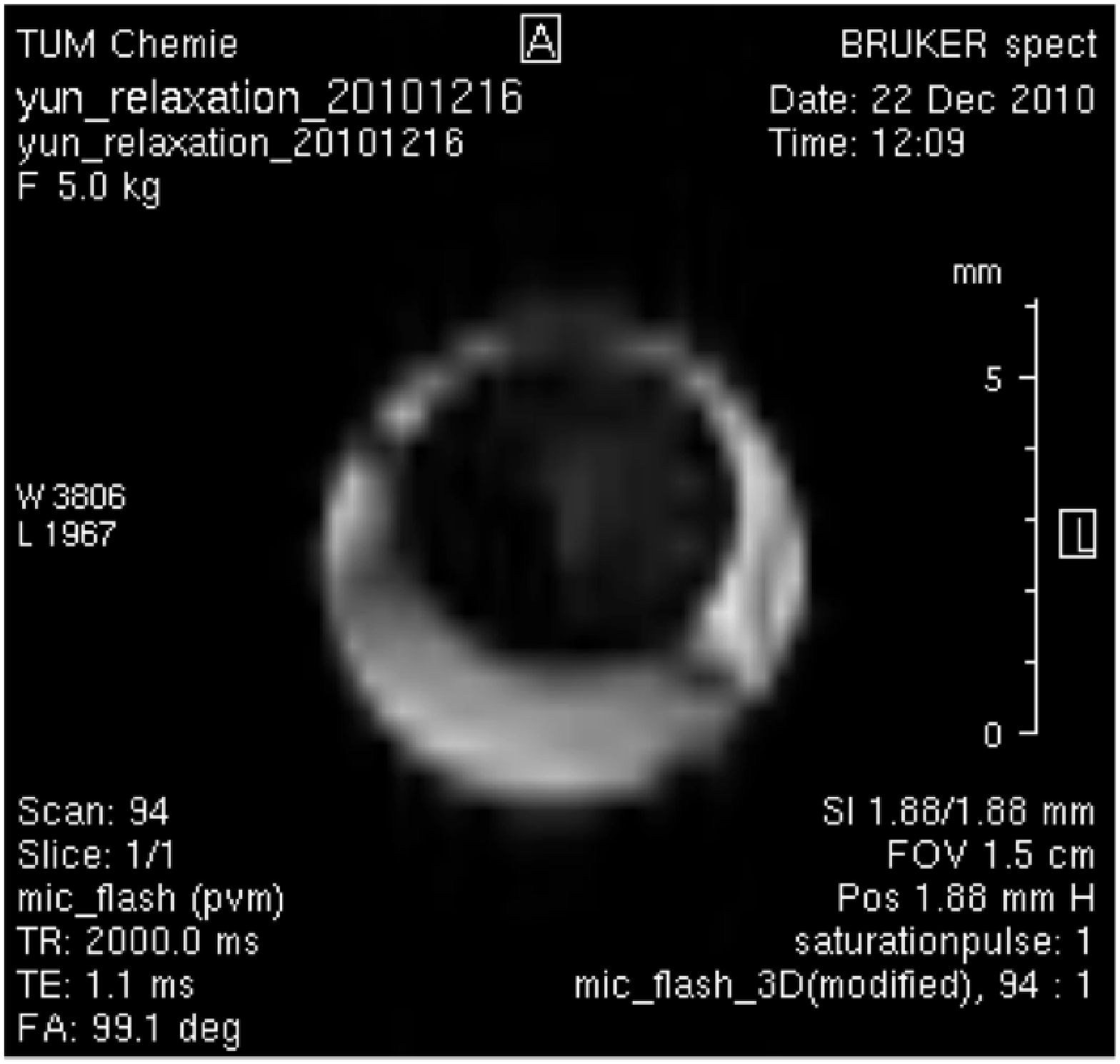}
\includegraphics[width=\sizefig\textwidth]{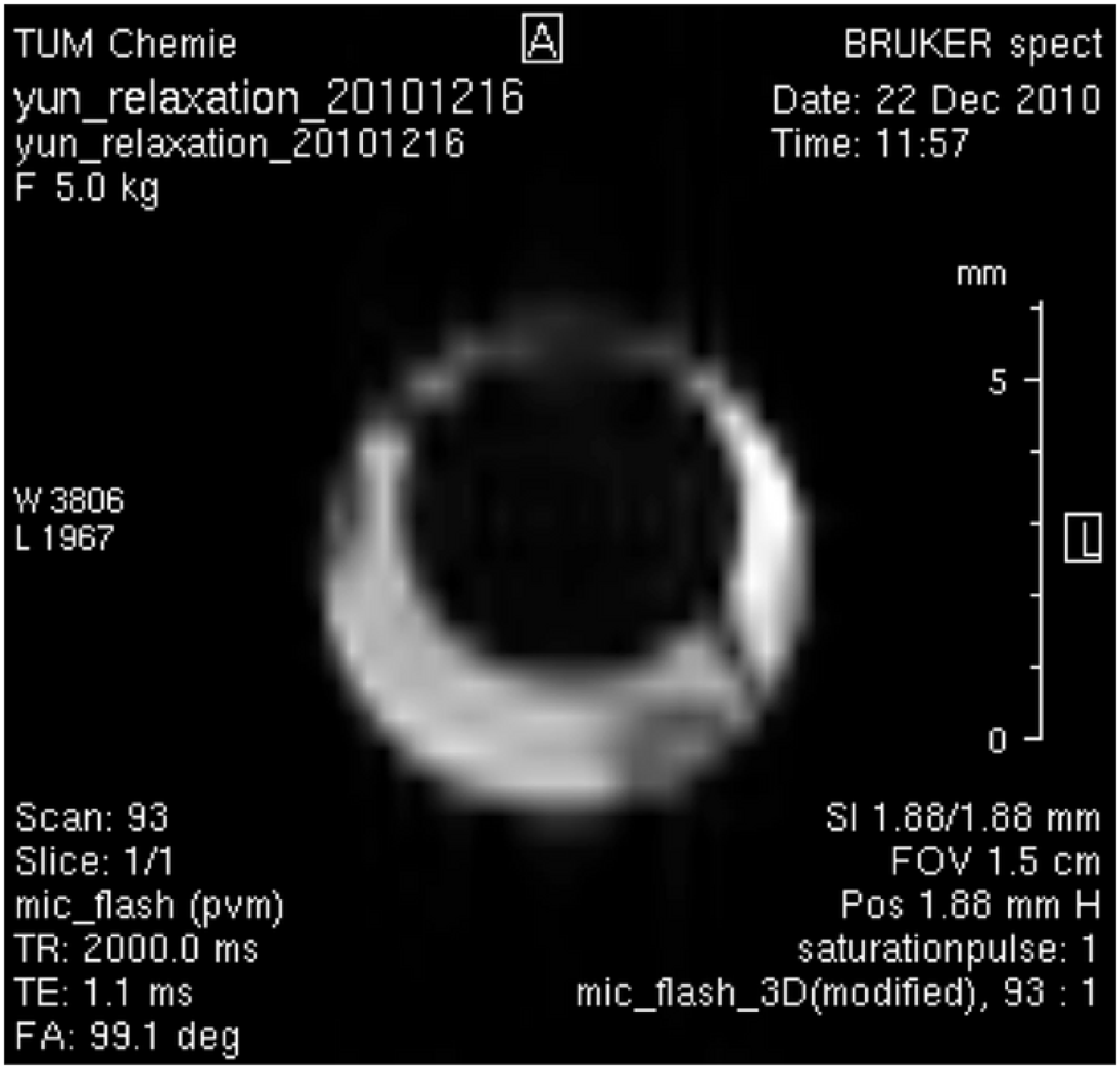}
\caption{\label{fig13}  Experimental results: The inner circle shape sample mimics the deoxygenated blood, where
$T_1=1.3$ s and $T_2=50$ ms; the outside moon shape sample corresponds to the oxygenated blood,
where $T_1=1.3$ s and $T_2=200$ ms. The goal of the control is to saturate the inner sample and to maximize the
remaining magnetization of the outside sample. The upper image is a reference image after a short
$90$ degree pulse on both samples. The image at the middle is the remaining Y magnetization $|M_y|$ after
the optimized pulse, the lower image is the remaining Z magnetization $|M_z$ after the optimized pulse.}
\end{figure}

\subsection{Some preliminary experimental results}
The first experimental results on the contrast problem are represented on Fig. \ref{fig13} and correspond to samples reproducing the case of the deoxygenated/oxygenated blood. Such results can be compared to Fig. \ref{fig11} where they have been numerically simulated for other samples in an ideal experiment. The preliminary experimental results are promising even if some artefacts due to the inhomogeneities of the magnetic field deteriorate the quality of the image.

\section{Conclusion}
In the conclusion, we discuss some important issues related to our study.
\paragraph*{Mathematical problems.} The important remaining question is to analyze the dynamics of the
singular flow in relation with the relaxation times and in particular the asymptotic of the trajectories.
\paragraph*{Numerical problems.} The main points consist of generating accurately complicated Bang-Singular
sequences solutions of the Maximum Principle and to prove the convergence of the continuation problems. In this setting, the problem is to initialize the shooting equation using the continuation method.
\paragraph*{Improving experimental results.} The preliminary experimental figure \ref{fig13} shows the problem
of the magnetic fields inhomogeneities, which have therefore to be taken into account in the model. In this case, the geometric techniques can be used as a first step to initialize a purely iterative numerical approach such as the GRAPE algorithm \cite{gershenzon,khaneja2,skinner,skinner2}. In this setting, the geometric solution provides an efficient initial solution and gives the physical limit of the contrast problem that can be reached. The GRAPE algorithm is then able to solve the simultaneous optimal control of a large number of spins of an inhomogeneous ensemble. In the example where the goal is to saturate the spins in deoxygenated blood, while maximizing the final magnetization of oxygenated blood, the numerically optimized pulse achieved about 93\% and 70\% of the contrast found by the optimal geometric solution for the ideal and the real cases. Note that this problem with magnetic field inhomogeneities is related to the controllability analysis of \cite{beauchard}.\\ \\
\noindent \textbf{Acknowledgment.}\\
B. B. and D. S acknowledge support from the PEPS INSIS
\emph{Optimal control of spin dynamics in Nuclear Magnetic
Resonance Imaging}.


\begin{thebibliography}{20}
\bibitem{altafini} \textsc{C. Altafini}, {\em Controllability properties
for finite dimensional quantum Markovian master equations}, J.
Math. Phys. \textbf{44}, 2357 (2002).

\bibitem{assemat} \textsc{E. Ass\'emat, M. Lapert, Y. Zhang, M. Braun, S. J. Glaser and
D. Sugny}, {\em Simultaneous time-optimal control of the inversion
of two spin 1/2 particles}, Phys. Rev. A, \textbf{82}, 013415
(2010).

\bibitem{beauchard} \textsc{K. Beauchard, J.-M. Coron and P. Rouchon}, {\em Controllability issues for continuous-spectrum systems and ensemble controllability of Bloch equations}, Comm. Math. Phys. \textbf{296}, 525 (2010)

\bibitem{bonnardchyba} \textsc{B. Bonnard and M. Chyba}, {\em Singular
trajectories and their role in control theory}, Math. and
Applications 40, Springer-Verlag, Berlin (2003)

\bibitem{BCS} \textsc{B. Bonnard, M. Chyba and D. Sugny}, {\em Time-minimal control of dissipative two-level
quantum systems: The generic case}, IEEE Transactions A. C.,
\textbf{54}, 2598 (2009).

\bibitem{energy} \textsc{B. Bonnard, O. Cots, N. Shcherbakova and
D. Sugny}, {\em The energy minimization problem for two-level
dissipative quantum systems}, J. Math. Phys., \textbf{51}, 092705
(2010)

\bibitem{bonnardsugny} \textsc{B. Bonnard and D. Sugny}, {\em Time-minimal control of dissipative two-level
quantum systems: The integrable case}, SIAM J. Control Optim.,
\textbf{48}, 1289 (2009).

\bibitem{contrast1} \textsc{G. M. Bydder, J. V. Hajnal and I. R.
Young}, {\em MRI: Use of the inversion recovery pulse sequence},
Clinical Radiology, \textbf{53}, 159 (1998)

\bibitem{contrast2} \textsc{M. Carl, M. Bydder, J. Du, A. Takahashi and E.
Han}, {\em Optimization of RF excitation to maximize signal and
$T_2$ contrast of tissues with rapid transverse relaxation},
Magnetic Resonance in Medicine, \textbf{64}, 481 (2010)

\bibitem{matter} \textsc{K. V. R. Chary and G. Govil}, {\em NMR in
biological systems, from molecules to human}, Focus on structural
biology, vol. 6, Springer (2008)

\bibitem{ernst} \textsc{R. R. Ernst}, \emph{Principles of Nuclear Magnetic Resonance in one and two dimensions} (International Series of Monographs on Chemistry, Oxford University Press, Oxford, 1990)

\bibitem{gershenzon} \textsc{N. I. Gershenzon, K. Kobzar, B. Luy, S. J. Glaser and T. E. Skinner}, \emph{Optimal control design of excitation pulses that accomodate relaxation}, J. Magn. Reson. \textbf{188}, 330 (2007)

\bibitem{gorini} \textsc{V. Gorini, A. Kossakowski and E. C. G. Sudarshan},
{\em Completely positive dynamical semigroups of $N$-level
systems}, J. Math. Phys., \textbf{17}, 821 (1976).

\bibitem{hampath} http://apo.enseeiht.fr/hampath

\bibitem{khaneja} \textsc{N. Khaneja, R. Brockett and S. J. Glaser}, {\em Time optimal control in spin systems},
Phys. Rev. A, \textbf{63}, 032308 (2001).

\bibitem{khaneja2} \textsc{N. Khaneja, T. Reiss, C. Kehlet, T. Schulte-Herbr\"uggen and S. J. Glaser}, {\em Optimal control of coupled spin dynamics: Design of NMR pulse sequences by gradient ascent algorithms},
J. Magn. Reson. \textbf{172}, 296 (2005).

\bibitem{krener} \textsc{A. J. Krener}, {\em The high order
maximal principle and its application to singular extremals}, SIAM
J. Control Optimization, \textbf{15}, 2, 256 (1977)

\bibitem{kupka} \textsc{I. Kupka}, {\em Geometric theory of
extremals in optimal control problems. I. The fold and Maxwell
case}, Trans. Amer. Math. Soc. \textbf{299}, 1, 225 (1987)

\bibitem{lapert} \textsc{M. Lapert, Y. Zhang, M. Braun, S. J. Glaser and D. Sugny}, {\em Singular extremals for the
time-optimal control of dissipative spin 1/2 particles}, Phys.
Rev. Lett., \textbf{104}, 083001 (2010)

\bibitem{spin} M. H. Levitt 2008 \emph{Spin dynamics: basics of nuclear magnetic resonance} (John Wiley and sons, New York-London-Sydney)

\bibitem{lindblad} \textsc{G. Lindblad}, {\em On the generators
of quantum dynamical semi-groups}, Comm. Math. Phys., \textbf{48},
119 (1976).

\bibitem{mirrahimi} \textsc{M. Mirrahimi and P. Rouchon}, {\em
Singular perturbations and Lindblad-Kossakowski differential
equations}, IEEE Trans. Automatic Control, \textbf{54}, 6, 1325
(2009)

\bibitem{pont} \textsc{L. Pontryagin et al}, {\em Th\'eorie
math\'ematique des processus optimaux}, Mir, Moscow, 1974.


\bibitem{skinner} \textsc{T. E. Skinner, T. O. Reiss, B. Luy, N. Khaneja and S. J.
Glaser}, {\em Application of Optimal Control Theory to the Design of Broadband Excitation Pulses for High Resolution NMR}, J. Magn. Reson. \textbf{163}, 8 (2003)

\bibitem{skinner2} \textsc{T. E. Skinner, T. O. Reiss, B. Luy, N. Khaneja and S. J.
Glaser}, {\em Tailoring the optimal control cost function to a desired output: Application to minimizing phase errors in short broadband excitation pulse}, J. Magn. Reson. \textbf{172}, 17 (2005)

\bibitem{sugnykontz} \textsc{D. Sugny, C. Kontz and H. R. Jauslin},
{\em Time-optimal control of a two-level dissipative quantum
system}, Phys. Rev. A, \textbf{76}, 023419 (2007).

\bibitem{viellard} \textsc{T. Viellard, F. Chaussard, D. Sugny, B.
Lavorel and O. Faucher}, {\em Field-free molecular alignment of
$CO_{2}$ mixtures in presence of collisional relaxation}, J. Raman
Spec., \textbf{39}, 694 (2008).

\bibitem{water} \textsc{C. Westbrook and C. Roth}, {\em MRI in
practice} (3rd Edition), Blackwell Publishing Ltd. (2005)

\bibitem{zhang} \textsc{Y. Zhang, M. Lapert, M. Braun, D. Sugny
and S. J. Glaser}, {\em Time-optimal control of spin 1/2
particules in presence of relaxation and radiation damping
effects}, J. Chem. Phys. \textbf{134}, 054103 (2011)
\end{thebibliography}
\end{document}